\documentclass[acmsmall,screen,nonacm]{acmart}

\usepackage{graphicx}
\usepackage{booktabs}
\newcommand{\Public}{\ensuremath{\mathsf{P}}}
\newcommand{\Transient}{\ensuremath{\mathsf{T}}}
\newcommand{\Secret}{\ensuremath{\mathsf{S}}}
\newcommand{\Low}{\mathsf{L}}
\newcommand{\High}{\mathsf{H}}
\newcommand{\Step}{\mathsf{step}}
\newcommand{\Force}{\mathsf{force}}
\newcommand{\Read}{\mathsf{read}}
\newcommand{\Write}{\mathsf{write}}
\newcommand{\Branch}{\mathsf{branch}}
\newcommand{\Ready}[1]{\mathsf{ready}(#1)}
\newcommand{\Pending}[3]{\mathsf{pending}(#1,#2,#3)}
\newcommand{\pubat}[2]{\mathsf{public}(#1,#2)}
\newcommand{\Traces}{\mathsf{Tr}}
\newcommand{\Harden}{\mathsf{Harden}}
\newcommand{\Beh}{\mathsf{Beh}}
\newcommand{\protectty}{\mathsf{protect}}
\newcommand{\releasety}{\mathsf{declassify}}
\newcommand{\execprefix}[2]{\mathrel{\overset{#2}{\underset{#1}{\Longrightarrow}}}}
\newtheorem{sslhTheorem}{Theorem}
\newtheorem{sslhDefinition}{Definition}

\title{VellSSLH: Sound Type-Guided Selective Speculative Load Hardening for LLVM IR}
\author{Swarn Priya}
\affiliation{%
  \institution{Washington State University}
  \country{USA}
}
\email{swarn.priya@wsu.edu}
\renewcommand{\shortauthors}{Swarn Priya}

\setcopyright{none}
\date{}
\hypersetup{
  pdftitle={VellSSLH: Sound Type-Guided Selective Speculative Load Hardening for LLVM IR},
  pdfauthor={Swarn Priya},
  pdfsubject={Selective speculative load hardening for LLVM IR}
}

\begin{document}
\raggedbottom
\setlength{\textfloatsep}{12pt plus 2pt minus 2pt}
\begin{abstract}
Constant-time cryptographic programs written in C can still leak secrets 
during speculative execution. Mispredicted branches may allow instructions 
to access secret data and expose it through observable behavior, even when 
the program is secure during normal execution. Speculative load hardening (SLH) 
addresses these leaks, but protecting values unnecessarily can introduce 
significant runtime overhead. We present VellSSLH, an approach that 
selectively inserts speculative load hardening at LLVM IR. Starting from 
annotated C programs, programmers specify Public and Secret inputs, and 
our analysis infers which values may become secret during speculation. 
The transformation then inserts misspeculation-flag operations and 
protections where needed, rather than protecting every potentially 
affected value. VellSCT, our extracted checker, validates the 
resulting LLVM IR.

We formalize speculative constant-time (SCT) security by extending 
Vellvm's LLVM semantics with speculative execution, attacker directives, 
and leakage observations. Our security type system reasons about both 
normal and speculative execution, including SSA joins, symbolic values, 
poison, pointer provenance, and observable faults. We also support explicitly 
declared declassifications and trusted external services through security 
policies and contracts. In Rocq, we prove that programs accepted by the 
checker satisfy SCT under the stated runtime assumptions. 
We separately prove that a reference hardening transformation preserves 
architectural behavior, ensuring that the inserted protections do not 
change the original program's defined behavior during normal execution. 
Evaluated optimized variants are rechecked for SCT and tested functionally, 
but are not covered by this architectural preservation proof. 
At the LLVM backend, we maintain a nonspillable misspeculation flag 
without permanently reserving a physical register. 
Our mechanized guarantees apply at LLVM IR; security of the generated binary 
additionally depends on backend and hardware assumptions.

We evaluate VellSSLH on 24 cryptographic API and component 
entries using a bare-metal Intel processor. Across nine layout-randomized trials,
VellSSLH achieves 4.03\% lower geometric-mean runtime than LLVM SLH. 
Thirteen comparisons clearly favor VellSSLH, eight favor LLVM SLH, and 
three remain inconclusive. The results also show that fixed entry barriers, 
register pressure, and reload protection can offset the benefits of 
removing unnecessary masks. For X25519, VellSSLH improves runtime by 22.69\% 
over LLVM SLH while introducing only 1.86\% overhead relative to unprotected code. 
Compared with full Ultimate SLH, VellSSLH achieves 19.65\% lower runtime, 
although this comparison also involves differences in LLVM versions and 
experimental campaigns. Functional testing and eight successful bounded 
binary-analysis checks provide additional evidence beyond the mechanized 
LLVM IR results.
\end{abstract}
\maketitle

\section{Introduction}
\label{sec:introduction}

Constant-time cryptography avoids secret-dependent branches and memory
addresses. A mispredicted branch can nevertheless bypass a bounds check and
access secret data, violating speculative constant-time (SCT)~\cite{spectector,jasmin-sslh}.
LLVM's speculative load hardening (SLH) tracks misspeculation with a predicate
and introduces dependencies to prevent leakage~\cite{llvm-slh}. Maintaining
that predicate increases register pressure; hardening also adds instructions
and dependencies. Selective hardening protects only values that can cause
speculative leakage.

\emph{Spectre Declassified}~\cite{spectre-declassified} introduced selective
SLH. Subsequent work on Jasmin~\cite{jasmin-sslh} distinguishes values that
are public in both normal and speculative execution, secret in both, or
normally public but potentially secret during misspeculation. We build on
this type system and its forward execution model. Only the third category
needs protection before observable uses; secret arithmetic need not be
masked. Blade~\cite{blade} similarly separates speculative sources from
observable sinks and protects paths between them.

Our goal is to bring mechanized guarantees to C-based cryptography through
LLVM. Ultimate SLH~\cite{ultimate-slh} implements hardening in LLVM with a
core-language security argument, but its LLVM transformation is not
mechanically verified. Jasmin provides verified selective hardening for its
own language rather than C. We use Vellvm~\cite{vellvm-modular} to account
for LLVM's SSA control flow, symbolic values, poison and memory semantics.
We extend it with speculative execution and observations, mechanize checker
soundness and reference-insertion preservation in Rocq, and extract the
analysis into LLVM.

\paragraph{Problem statement and contributions}
We present VellSSLH, an approach to selectively hardening constant-time C
implementations against Spectre v1 at LLVM IR. Programmers specify which
inputs are public or secret; the compiler infers and inserts the speculative
protections. Section~\ref{sec:llvm-challenges} explains the LLVM-specific difficulties.
Our contributions are as follows:
\begin{itemize}
\item We define an SCT type system called VellSCT for LLVM IR with normal/speculative
security types and MSF typestate, handling PHIs, poison, under-defined values
and pointer operations. We extend Vellvm with speculative execution and
leakage observations, and prove checker soundness in Rocq, including explicit
declassification and trusted external services.
Our Rocq development adds 337,040 lines of code and proofs, excluding
existing Vellvm sources, generated files, comments and blank lines.
\item We develop automatic analysis and hardening from Public/Secret
annotations, generating MSF initialization, edge updates and protections.
We prove architectural preservation for a reference insertion algorithm in
Rocq, including normal results, memory effects and faults. Evaluated optimized
variants are rechecked by VellSCT and tested functionally; this preservation
proof does not cover those variants.
\item We implement VellSSLH in LLVM using the extracted VellSCT checker. 
Our backend handles compiler-generated spills and ensures that the misspeculation 
flag (MSF) is never spilled, without permanently reserving a physical register. 
We evaluate the x86-64 backend and distinguish the mechanized IR guarantees
from the conditional backend argument.
\item We evaluate 24 cryptographic API/component entries spanning symmetric
primitives, hashes, X25519 and Kyber Round~3. On bare-metal Intel, VellSSLH
reduces geometric-mean runtime by 4.03\% against LLVM SLH and 19.65\% against
full Ultimate SLH, with version and campaign effects disclosed. Functional
tests and bounded binary analysis complement timings; a controlled three-core
experiment isolates selectivity, reducing runtime by 2.3--5.3\% relative to
conservative placement.
\end{itemize}

\begin{figure}[!tbp]
\centering
\includegraphics[width=\linewidth]{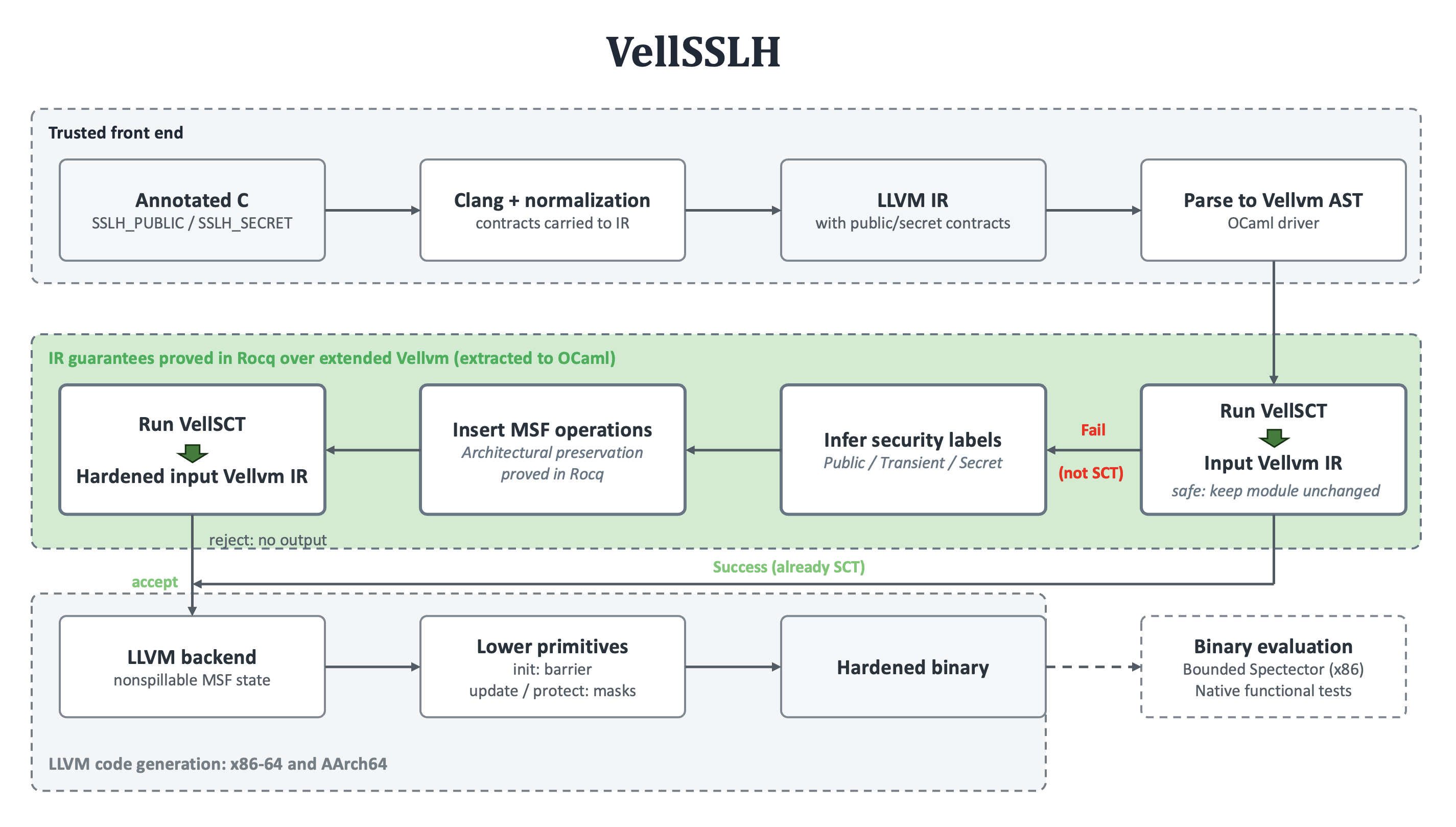}
\caption{The VellSSLH workflow.}
\label{fig:workflow}
\Description{Annotated C passes through Clang, normalized LLVM IR and the Vellvm parser. 
VellSCT either accepts the input unchanged or the hardener infers labels, inserts MSF operations 
and checks the candidate. Only accepted IR proceeds to backend lowering and binary evaluation.}
\end{figure}
\paragraph{Overview of VellSSLH}
In Figure~\ref{fig:workflow}, Clang carries C contracts into normalized LLVM
IR, which an OCaml driver parses into Vellvm's AST. VellSCT returns accepted
inputs unchanged. A failed check leaves SCT unestablished; the figure's
``not SCT'' branch has this meaning. For uncertified inputs, inference guides MSF initialization, branch
updates, and protections, followed by another check; failure rejects the
transformation. Declassification and external functions use explicit
policies and contracts. Accepted IR is compiled with a nonspillable MSF;
Section~\ref{sec:implementation} describes backend checks and binary testing.

\section{Motivating Example}
\label{sec:motivation}

Figure~\ref{fig:example} illustrates selective hardening
using two LLVM IR functions. Both load a value after a
public bounds check. Because \texttt{\%p} points to public
data, the loaded value is public during normal execution.
However, a mispredicted branch may allow the load to
access secret data, making the value \emph{Transient}. 
In \texttt{lookup}, the loaded value \texttt{\%v}
reaches a public return. VellSSLH inserts a
\texttt{protect} operation that preserves the value
during normal execution and replaces it with a fixed
public value during misspeculation.

\begin{figure}[!htb]
\centering
\begin{minipage}[t]{0.48\linewidth}
\footnotesize
\textbf{(a) Protection required}
\begin{verbatim}
; p: public data; i/n/return: public
; primitive calls abbreviated
define i32 @lookup(ptr %p,
                   i64 %i, i64 %n) {
entry:
  %m0 = init_msf()
  %p0 = protect(%p, %m0)
  %i0 = protect(%i, %m0)
  %n0 = protect(%n, %m0)
  %c = icmp ult i64 %i0, %n0
  %g = freeze i1 %c
  br i1 %g, label %yes, label %no
yes:
  %m1 = set_msf(%m0, %g, true)
  %a = getelementptr i32, ptr %p0,
                          i64 %i0
  %v = load i32, ptr %a
  %r = protect(%v, %m1)
  ret i32 %r
no:
  %m2 = set_msf(%m0, %g, false)
  ret i32 0
}
\end{verbatim}
\end{minipage}\hfill
\begin{minipage}[t]{0.48\linewidth}
\footnotesize
\textbf{(b) No payload protection needed}
\begin{verbatim}
; p: public data, out: secret data
; i/n: public
define void @compute(ptr %p,
                     ptr %out,
                     i64 %i, i64 %n) {
entry:
  %c = icmp ult i64 %i, %n
  br i1 %c, label %yes, label %no
yes:
  %a = getelementptr i32, ptr %p, i64 %i
  %v = load i32, ptr %a
  %x = add i32 %v, 1
  store i32 %x, ptr %out
  ; no protect needed for %v
  ret void
no:
  ret void
}
\end{verbatim}
\end{minipage}
\caption{Hardened Public-return example (a); source fragment with a Secret store (b).}
\label{fig:example}
\Description{Two LLVM IR functions showing how
VellSSLH selectively protects Transient values.}
\end{figure}

In \texttt{compute}, arithmetic and a store to Secret memory at a Public
address need no payload protection for \texttt{\%v}. Entry/address hardening
is omitted in panel (b). Panel (a) shows all flag operations, the shared
frozen guard and entry views before pruning.

\subsection{Challenges of Selective Hardening in LLVM}
\label{sec:llvm-challenges}
The security lattice alone does not account for LLVM's SSA representation,
value semantics or automatic register allocation.

\paragraph{Flag flow at SSA joins.}
The flag must follow the incoming edge, including loop backedges:
\begin{quote}\small
\begin{minipage}{\linewidth}
\begin{verbatim}
left:  mL = set_msf(m,c,true);  br join
right: mR = set_msf(m,c,false); br join
join:  mJ = phi [mL,left], [0,right]  ; incorrect
       r  = protect(x,mJ)
\end{verbatim}
\end{minipage}
\end{quote}
If $c$ is true but speculation takes the right edge, $m_R=-1$.
The PHI nevertheless selects zero and loses the protection. Replacing
\texttt{[0,right]} with \texttt{[mR,right]} fixes it. LLVM's verifier accepts
both integer PHIs; VellSCT additionally checks predecessor-specific flag
flow. The proof relates the saved predecessor environments, not sequential
updates of PHI destinations.

\paragraph{Poison and unresolved choices.}
LLVM values can be \emph{poison}, a placeholder for an invalid result.
Most operations simply propagate poison: if either operand of an
\texttt{or} is poison, the result is poison. This complicates implementing
protection at the IR level. During misspeculation, integer \texttt{protect}
must replace its input with a fixed value ($-1$), whatever that input
is, including poison. If \texttt{protect} were an ordinary \texttt{or},
a poison input would stay poison instead of becoming the fixed value. We
therefore give \texttt{protect} its own semantics, in which a set flag
always produces $-1$. The Rocq security proof uses these semantics;
machine lowering requires a separate backend argument. Poison also affects
branches: branching on poison is undefined behavior. We use \texttt{freeze}
to choose one concrete Boolean, and the checker ensures that the branch
and its flag update both use that same frozen Public value.

\paragraph{Pointer identity and memory effects.}
Equal pointer bits need not grant equal access: permissions and allocation
provenance also matter. Clean protection preserves the original pointer;
set protection may construct a different pointer or fail. Both the security
and architectural proofs account for these outcomes, rather than treating
pointers as mathematical integers.

\paragraph{Automatic spills.}
Jasmin makes register and stack storage explicit~\cite{jasmin-2017}; LLVM
introduces spills after IR checking. A Public value reloaded during
misspeculation may be Transient. Worse, reloading a clean MSF would disable
later protections. We tag the MSF as nonspillable through copies, PHIs and
allocation. LLVM chooses its physical register; none is permanently reserved.
Allocation rejects if this constraint cannot be met. At calls the active
flag ends at a synchronization boundary, avoiding a new calling convention.
Ordinary spills remain allowed, with late checks certifying Public reloads
or protecting demanded Transient reloads. Other backend obligations remain
conditional (Section~\ref{sec:implementation}).

\section{LLVM with Speculative Observations}
\label{sec:model}
We model Spectre v1 by extending Vellvm's denotations, event interface,
and handlers~\cite{vellvm-modular,vellvm-experience} with attacker-directed
forward speculation and leakage observations, reusing its AST and value
domains. The attacker observes branch decisions, memory addresses, public
results, and authorized disclosures.

\subsection{Syntax and LLVM values}

\begin{figure}[!tbp]
\centering\small
\begin{minipage}[t]{0.48\linewidth}
\vspace{0pt}
\(\displaystyle
\begin{array}{@{}rcl@{}}
 \kappa &::=& \mathtt{i}w\ (1\leq w\leq64)
       \mid\mathtt{ptr}\mid\mathtt{void}\\
 e &::=& n\mid\%x\mid @g\mid\mathtt{null}\\
 op &::=& e_1\mathbin{\diamond}e_2
       \mid\mathtt{icmp}\ e_1,e_2
       \mid\mathtt{cast}\ e\\
    &\mid&\mathtt{gep}\ T,e,\vec e
       \mid\mathtt{select}\ e,e_1,e_2\\
 \phi &::=& x=\Phi[\ell_1:e_1,\ldots,\ell_n:e_n]\\
 term &::=&\mathtt{br}\ e,\ell_1,\ell_2
       \mid\mathtt{br}\ \ell\\
      &\mid&\mathtt{ret}\ e
       \mid\mathtt{ret}\ \mathtt{void}\\
 block &::=&\ell:\ \vec\phi;\vec i;term\\
 cfg &::=&\mathtt{define}\ \kappa\ f(\vec x)\
           \{\vec{block}\}
\end{array}
\)
\end{minipage}\hfill
\begin{minipage}[t]{0.50\linewidth}
\vspace{0pt}
\(\displaystyle
\begin{array}{@{}rcl@{}}
 i &::=& x=op\mid x=\mathtt{alloca}\ T\\
   &\mid& x=\mathtt{load}\ \mathtt{i}w,e\\
   &\mid&\mathtt{store}\ e_1,e_2
       \mid\mathtt{call}\ \mathtt{void}\ f(\vec e)\\
   &\mid&x=\mathtt{freeze}\ \mathtt{i1}\ e\\
   &\mid&m=\mathtt{init\_msf}()\\
   &\mid&m'=\mathtt{set\_msf}(m,c,t)\\
   &\mid&x=\mathtt{protect}_{\kappa}(e,m)\\
   &\mid&x=\mathtt{declassify}_{s}(e)\\
 P &::=&\vec{cfg},\vec g
\end{array}
\)
\end{minipage}
\vspace{-1ex}
\caption{Syntax of the checked LLVM IR fragment; $T$ includes array and structure layouts.}
\label{fig:syntax}
\Description{Grammar of scalar SSA functions, blocks, PHIs, terminators, memory instructions, and explicit hardening and declassification operations.}
\end{figure}

Figure~\ref{fig:syntax} gives the checked fragment. A module contains
functions and supported globals; each function is a CFG of blocks
containing PHIs, instructions, and a terminator.

The SSLH operations \texttt{init\_msf}, \texttt{set\_msf}, and
\texttt{protect} abbreviate calls to \texttt{llvm.sslh.init.msf},
\texttt{llvm.sslh.update.msf}, and \texttt{llvm.sslh.protect.*},
represented by Vellvm's existing call AST and interpreted
by our extended semantics. Declassification abbreviates
a marker call authorized by an explicit policy.

Loads, stores, and returns support integer values; returns may also be
void. Ordinary calls return void. Pointer-valued loads, stores, and returns
are unsupported. The checker validates CFG edges, incoming PHI operands,
SSA definitions, and dominance.

Vellvm distinguishes dynamic values (\emph{dvalues}) $dv\in V$---integers,
pointers, and poison---from under-defined values (\emph{uvalues}) $uv\in V_u$,
which can also retain symbolic expressions and unresolved choices.
Concretization resolves choices when an operation requires them; a dvalue
need not be defined in LLVM's sense. The memory model tracks allocations,
permissions, and provenance, so equal address bits need not imply equal
memory behavior.

\subsection{Speculative Execution Semantics}
\label{sec:spec-semantics}

An interaction tree can return a result, take an internal step, or perform
an event and continue. We add branch-selection and protection events,
reusing Vellvm's memory, environment, stack, and value operations.
A configuration $C=\langle t,\sigma\rangle$ pairs the remaining computation
$t$ with state $\sigma=(H,\rho,F,b)$: memory $H$, local environment $\rho$,
saved environments $F$, and semantic misspeculation bit $b$. This bit
records actual misspeculation; the program separately maintains the MSF.

\paragraph{Attacker control and observations.}
The attacker forces incorrect branches and redirects unsafe speculative
memory accesses. Execution follows directives $d$ and produces
observations $o$. We write
$C\xrightarrow[d]{o}C'$ for a transition from
configuration $C$ to $C'$ under directive $d$,
producing observation $o$:
\[
\begin{array}{rcl}
d &::=& \Step \mid \Force
       \mid \mathsf{load}(a)
       \mid \mathsf{store}(a),\\
o &::=& \epsilon \mid \Read(a)
       \mid \Write(a) \mid \Branch(c)
       \mid \mathsf{return}(v?)
       \mid \mathsf{fault}(r)
       \mid \mathsf{declassify}(s,q).
\end{array}
\]
Here, $a$ is a Vellvm memory address. Return values are
observable only for public root returns. At an authorized
site $s$, declassification reveals the materialized outcome
$q$: a value or a failure of concretization.
Fault observations include undefined behavior,
out-of-memory, unsupported operations, and blocked
speculation barriers.

\paragraph{Speculative branches.}
At a conditional branch, $\Step$ follows the evaluated
condition $c$, while $\Force$ follows the opposite branch.
A forced branch sets $b$ to true, indicating
misspeculation. Once set, $b$ remains true throughout
the remaining execution. Both choices produce the
observation $\Branch(c)$, which records the evaluated
condition rather than the branch selected by the attacker.

\paragraph{Speculative memory accesses.}
During normal execution, loads and stores use their
requested addresses. During misspeculation, if a load
accesses an unreadable region, the attacker may redirect
it to another readable address. Stores follow a similar
rule when their requested address is not writable.
The observation records the original requested address,
not the redirected address or the accessed value.
A valid memory access cannot be redirected simply because
execution is speculative. However, an invalid speculative
load may be redirected to a secret memory region, allowing
a normally public load to return secret data. We therefore
assign such a loaded value the Transient type. Readability and writability
include provenance checks: landing in a different allocated object is not
enough if the pointer lacks permission for that allocation. A pointer whose
provenance does permit that writable access is not redirected. This is the
chosen forward-speculation contract, not a model of all speculative stores.

\begin{figure}[!tbp]
\centering\small
\(\displaystyle
\begin{gathered}
\frac{\ }
{\langle\mathsf{choose}(c,k),(H,\rho,F,b)\rangle
 \xrightarrow[\Force]{\Branch(c)}
 \langle k(\neg c),(H,\rho,F,\mathsf{true})\rangle}
\quad\text{\textsc{Branch-Force}}
\\[0.5ex]
\frac{
  \neg\mathsf{readable}_T(H,a)
  \quad
  \mathsf{read}_T(H,a')\Downarrow uv
}{
  \langle\mathsf{load}(T,a,k),(H,\rho,F,\mathsf{true})\rangle
  \xrightarrow[\mathsf{load}(a')]{\Read(a)}
  \langle k(uv),(H,\rho,F,\mathsf{true})\rangle
}
\quad\text{\textsc{Load-Redirect}}
\\[0.5ex]
\frac{b=\mathsf{false}}
{\langle\mathsf{init}(k),(H,\rho,F,b)\rangle
 \xrightarrow[\Step]{\epsilon}
 \langle k(0),(H,\rho,F,b)\rangle}
\quad\text{\textsc{Init}}
\\[0.5ex]
\frac{b=\mathsf{true}}
{\langle\mathsf{init}(k),(H,\rho,F,b)\rangle
 \xrightarrow[d]{\mathsf{fault}(\mathsf{barrier})}
 \mathsf{stopped}(H,\rho,F,b)}
\quad\text{\textsc{Init-Blocked}}
\end{gathered}
\)
\caption{Selected speculative execution rules; $k$ is the continuation.}
\label{fig:directed-semantics}
\Description{Formal transition rules for attacker-directed
speculative execution, memory redirection, and MSF
initialization.}
\end{figure}

Figure~\ref{fig:directed-semantics} gives representative event rules after
operand evaluation and required concretization. Initialization sets the
MSF to zero during normal execution and stops misspeculating execution,
preventing a reset. Speculation proceeds forward without rollback
(Section~\ref{sec:limitations}). The barrier fault represents a stopped
speculative window, rather than an architectural trap. It is observable in
both related executions because their semantic bits agree.

\subsection{Protection primitives}
The program uses a misspeculation flag $m$, encoded
as $0$ for clean and $-1$ (all ones) for set. The
checker enforces flag-flow constraints that support
the invariant $b=\mathsf{true}\Rightarrow m=-1$:
whenever execution is misspeculating, the current
flag must be set. This implication suffices for protection;
it does not require a set flag to imply misspeculation.

The SSLH operations behave as follows, where
$c$ is the branch condition and $t$ is the branch
taken:
\[
\begin{gathered}
\mathtt{init\_msf}() = 0 \quad \text{if } b=\mathsf{false},\\[-2pt]
\begin{aligned}
\mathtt{set\_msf}(m,c,t)
&= m\mathbin{\mathrm{OR}}
\begin{cases}
0 & \text{if } c=t,\\
-1 & \text{otherwise},
\end{cases}
\qquad
\mathtt{protect}_{w}(v,m)
&= (v\mathbin{\mathrm{OR}}m)\bmod 2^w.
\end{aligned}
\end{gathered}
\]

These equations apply to defined integer payloads, a canonical 64-bit
MSF, and defined Boolean update operands. A mismatched edge sets the flag
permanently across updates. Protection with a clean flag preserves the
original LLVM value, including poison, under-defined expressions, and
pointer provenance. With a set flag, it ignores the payload and produces
all ones (Section~\ref{sec:llvm-challenges}). For pointers it invokes
Vellvm's integer-to-pointer operation on $2^{64}-1$ with wildcard
provenance; this can report out-of-memory, and success does not imply
dereferenceability.

\section{Contract-Guided Security Typing and Hardening}
\label{sec:types}

Source contracts classify normal-execution data; inference extends these
classifications to speculation, guides insertion, and checks hardened IR.

\subsection{Contracts and Security Types}
\label{sec:contract-types}
Scalar inputs use \texttt{SSLH\_PUBLIC} or \texttt{SSLH\_SECRET}; return
contracts classify outputs. Pointer contracts distinguish addresses
from memory contents: for example,
\texttt{SSLH\_SECRET\_DATA} marks the memory contents as
secret, while the pointer itself may remain public.
Callers must satisfy these contracts. Writes may change the inferred secrecy
of memory contents. The checker validates instruction/SSA shapes, not general
bounds, initialization or separation proofs. Recursive allocation-layout
validity and admissible initial memory are theorem premises. Vellvm still
models access failures; acceptance alone is not a memory-safety theorem.

We assign each value a security type that describes its
behavior during normal and speculative execution. Let
$\Low \leq \High$, where $\Low$ means public and $\High$
means secret. A security type is a pair
$\tau=(\tau_n,\tau_s)$, with $\tau_n\leq\tau_s$.
This gives three security types:
\[
\Public=(\Low,\Low),\qquad
\Transient=(\Low,\High),\qquad
\Secret=(\High,\High).
\]

A \Public{} value is public in both executions.
A \Transient{} value is public during normal execution
but may become secret during speculation. A \Secret{}
value is secret in both executions. The types form the
ordering 
$
\Public\leq\Transient\leq\Secret,
$
with join operator $\sqcup$.

The semantic bit $b$ determines which component of the
security type is used. We write $\pubat{\tau}{b}$ for
the normal component when $b$ is false and the speculative
component when $b$ is true. Two scalar values are related
at type $\tau$ when they must be equal whenever the
selected component is public:
$
 v_1\mathrel{\approx^b_\tau}v_2
 \Longleftrightarrow
 \pubat{\tau}{b}=\Low\ \Longrightarrow\ v_1=v_2.
$

Its extension to LLVM values and memory accounts for valid accesses,
under-defined expressions, and memory behavior.

The typing environment $\Gamma$ assigns each SSA variable
an LLVM type $\kappa$ and a security type $\tau$.
For each tracked pointer $q$, it also records the memory-content
security type $\mu_q$. We write $\operatorname{Ptr}(\Gamma)$ for
the tracked pointer bindings. Each basic block $\ell$ has a
control-flow security type $pc_\ell$. Integer and Boolean constants 
and null pointers are
initially Public. Raw \texttt{undef} and \texttt{poison}
operands are rejected. Public source annotations specify
that inputs are public during normal execution, but do
not automatically make them public during speculation.
Secret inputs remain Secret.

During inference, the security type determines whether an unsafe use can be made safe through hardening.
A transient value may be protected before a public observation, whereas
a Secret value cannot be repaired this way. Architectural constant-time is
an applicability requirement: insertion rejects unrepairable Secret observable
uses, and VellSCT checks the final Public-use constraints. Soundness does not
assume acceptance by a separate constant-time checker.

\subsection{SSA and Memory Constraints}

We write $\Gamma;pc\vdash e:\tau$ to mean that expression
$e$ has security type $\tau$ under environment $\Gamma$
and control type $pc$. For instructions with memory
effects, $\Gamma;pc\vdash i\dashv C$ produces a set of
constraints $C$.

Arithmetic joins operand and control types. PHIs join incoming types and
the block's control type; pointer results also track memory contents.
Branches propagate guard types to successors. A fixed point over the CFG
jointly computes PHIs, control types, and memory constraints, including
loop backedges.

Figure~\ref{fig:type-rules} shows representative rules for checking
protected IR. Planning uses the normal-component criterion described below.

\begin{figure}[!tbp]
\centering\small
\(\displaystyle
\begin{gathered}
\frac{\Gamma;pc\vdash e_1:\tau_1\quad
      \Gamma;pc\vdash e_2:\tau_2}
 {\Gamma;pc\vdash
  e_1\mathbin{\diamond}e_2:
  pc\sqcup\tau_1\sqcup\tau_2}
 \quad\text{\textsc{Total-Op}}
\qquad
\frac{
  \Gamma(p)=(\mathtt{ptr},\tau_p,\mu_p)
  \quad pc\sqcup\tau_p\leq\Public
}{
  \Gamma;pc\vdash
  \mathtt{load}\ w,p:
  pc\sqcup\tau_p\sqcup\mu_p\sqcup\Transient
}
\quad\text{\textsc{Load}}
\\[0.5ex]
\frac{
  \Gamma(p)=(\mathtt{ptr},\tau_p,\mu_p)
  \quad \Gamma;pc\vdash v:\tau_v
  \quad pc\sqcup\tau_p\leq\Public
}{
  \Gamma;pc\vdash
  \mathtt{store}\ v,p\ \dashv
  \{pc\sqcup\tau_p\sqcup\tau_v\leq\mu_q
    \mid q\in\operatorname{Ptr}(\Gamma)\}
}
\quad\text{\textsc{Store}}
\\[0.5ex]
\frac{
  \Gamma;pc\vdash v:\tau
  \quad\Sigma=\Ready{m}
}{
  \Sigma;\Gamma;pc\vdash
  \mathtt{protect}(v,m):pc\sqcup\protectty(\tau)
}
\quad\text{\textsc{Protect}}
\qquad
\frac{
  \Gamma;pc\vdash c:\tau_c
  \quad pc\sqcup\tau_c\leq\Public
}{
  \Gamma;pc\vdash
  \mathtt{br}\ c,\ell_t,\ell_f\;\mathsf{ok}
}
\quad\text{\textsc{Branch}}
\end{gathered}
\)
\caption{Selected typing rules for protected LLVM IR.}
\label{fig:type-rules}
\Description{Typing rules for arithmetic, loads, stores,
protection, and conditional branches.}
\end{figure}
\textsc{Load} and \textsc{Store} require Public addresses. Loads are at
least Transient because an invalid speculative access can read Secret
memory. Stores constrain every tracked pointer's memory label to account
for aliasing; the extension also tracks global-memory facts. Solving these
constraints flow-insensitively means a store
can also raise an earlier load's label. This conservative treatment avoids
aliasing assumptions that may fail during speculation.

At LLVM IR, fixed-size allocations produce abstract addresses whose
security label is $pc$, hence Public in a Public control context.
Their initial contents are conservatively classified Secret.
Concrete stack layout is determined later; separate-allocation tracking
could improve precision.

The \textsc{Protect} rule requires a ready, current MSF and
joins the control label with the payload type transformed by
$\protectty$. The \textsc{Branch} rule requires the
branch condition and its control context to be Public. Thus every reachable
checked block has Public control; non-Public control labels matter during
planning, when they expose paths that require repair or rejection.

Partial operations can fail before their results are used, so relevant
operands of division, GEP, and other supported partial operations must
be Public. Secret \texttt{select} guards are rejected even if a target
could implement the selection without branching.

\subsection{Flag Typestate}

Protection must use the current flag. Its typestate is
$\Sigma::=\mathsf{unknown}\mid\Ready{m}\mid\Pending{m}{c}{t}$.

$\mathsf{unknown}$ means that no valid MSF is available.
$\Ready{m}$ identifies the current flag and requires $b\Rightarrow m=-1$.
$\Pending{m}{c}{t}$ records that edge $t$ of branch condition $c$ was
selected but its update is outstanding. Protection requires a ready flag;
an update must use the current flag, guard, and actual edge, making the
result ready again.

LLVM branch conditions may be poison or undefined.
We use a restricted \texttt{freeze i1} instruction
so that the branch and its MSF update use the same
defined Boolean. The checker verifies this exact
pattern; arbitrary uses of \texttt{freeze} are not
accepted through this exception.

Flag PHIs must select the current MSF from every incoming edge,
including loop backedges, preventing stale flags and resets to zero.

The main typestate transitions are:
\[
\begin{gathered}
 \mathsf{unknown}
 \xrightarrow{m=\mathtt{init\_msf}()}
 \Ready m,
 \qquad
 \Ready m
 \xrightarrow{\mathtt{br}\ c\ \mathrm{to}\ t}
 \Pending m c t,
 \\[0.5ex]
 \Pending m c t
 \xrightarrow{m'=\mathtt{set\_msf}(m,c,t)}
 \Ready{m'},
 \qquad
 \frac{
   \forall p\in\mathsf{pred}(\ell).
   \ e_p=\mathsf{out}(p)
 }{
   m=\Phi[p:e_p]_{p\in\mathsf{pred}(\ell)}
   \ \mathsf{valid}
 }.
\end{gathered}
\]

Here $\mathsf{out}(p)$ is predecessor $p$'s current MSF.
The checker restricts initialization and update placement in the CFG
(Section~\ref{sec:llvm-challenges}). MSF values cannot be stored,
returned, passed as ordinary arguments, or used in unrelated arithmetic.

\subsection{Protecting Transient Uses}

Protection is needed when a Transient value reaches a Public use:
$\protectty(\Public)=\Public$, $\protectty(\Transient)=\Public$,
and $\protectty(\Secret)=\Secret$.

Masking repairs speculative secrecy, so Transient becomes Public, whereas
Secret remains Secret. Planning allows an observable operand whose normal
component is public; checking requires both components to be public:
\[
\begin{aligned}
\Gamma;pc\vdash_{\mathsf{plan}}e\ \mathsf{usable}
&\quad\Longleftrightarrow\quad (pc\sqcup\Gamma(e))_n=\Low,\\
\Gamma;pc\vdash_{\mathsf{checked}}e\ \mathsf{usable}
&\quad\Longleftrightarrow\quad pc\sqcup\Gamma(e)=\Public.
\end{aligned}
\]

Thus Transient operands can be repaired, but Secret observable operands
are rejected. Secret additions, shifts, and bitwise operations need no
mask when they do not reach an observation; a Transient address does.

\subsection{Insertion, Checking, and Explicit
Declassification}

VellSCT first checks the input, returning accepted programs unchanged.
Otherwise, inference guides insertion of MSF initializations at required
entries, updates along conditional branches, flag PHIs at joins, and
protections before observable uses.

The insertion algorithm reuses already protected
Public values when possible. For ordinary SSA uses,
the protected definition must dominate the use.
For PHI nodes, the protected value must be available
along the corresponding incoming edge.

After insertion, inference is recomputed under the declassification policy
and masks of already-Public inputs are removed. VellSCT independently
checks the result's types and MSF constraints; failed checks and unsupported
cases reject the transformation. Placement is heuristic: we claim no minimum
number of masks or optimality relative to Blade's min-cut formulation.

\paragraph{Declassification.}
Declassification lets a program intentionally reveal a Secret value.
For example, Secretbox open reveals whether authentication succeeded. 
A declassification policy authorizes the value and type disclosed at each
site, such as an authentication result. The checker also requires
$pc=\Public$. For concrete scalars the value is preserved and its normal
component becomes public:
\[
\releasety(\Public)=\Public,\qquad
\releasety(\Transient)=\Transient,\qquad
\releasety(\Secret)=\Transient.
\]
However, declassification makes the value Public only during normal
execution. During misspeculation, the program may be running with wrong
values, so the result can still depend on secrets. We therefore give a
declassified value type Transient, not Public. If the program later
branches on the authentication result, VellSSLH still protects it
before the branch. In normal execution, declassification reveals the 
value to the attacker. During misspeculation, it reveals nothing. 
VellSSLH checks that values are declassified only at authorized sites, 
but it does not decide whether revealing them is safe; the programmer 
who writes the policy must justify each release.

\section{Formal Verification}
\label{sec:proofs}

In extended Vellvm, we prove VellSCT soundness, including declassification
and trusted services, and architectural preservation of the reference
inserter, under the stated program checks and runtime assumptions.

\subsection{Speculative constant-time}

An attacker strategy $A:\mathit{Obs}^*\to\mathit{Dir}$ chooses
the next directive based on previous observations. A fixed
directive sequence makes these choices in advance; a strategy
also permits adaptive choices. Both executions use the same
strategy, so equal observation histories receive the same
next directive. Silent steps do not produce observations. Let
$\Traces_A(P,\sigma)$ denote all possible finite observable
prefixes of program $P$ starting from state $\sigma$ under
strategy $A$. Prefixes may have any finite length, including
prefixes of nonterminating executions.

We adapt SCT to LLVM's nondeterministic semantics.
We compare sets of possible observation prefixes because allocation and concretization
may produce different results without leaking secrets.
Requiring observations to agree for every pair of independently chosen
executions would reject harmless nondeterminism. Instead, each execution
must have a counterpart with the same observations.

Let $\sigma_1\approx_I\sigma_2$ denote two initial states
related by the entry contract $I$. Both executions start
outside speculation, with empty local environments and
frame stacks. Public arguments must be the same, while Secret
arguments may differ. Their heaps must agree on the
declared public memory and have compatible allocations
and permissions. Public pointer contracts also cover
accessible memory bytes through pointer provenance.
Initial values and memory must satisfy Vellvm's
admissibility requirements.

\begin{sslhDefinition}[Speculative constant-time (SCT)]
A program $P$ is SCT with respect to $I$ if, for indistinguishable worlds,
$
 \sigma_1\approx_I\sigma_2
 \Longrightarrow
 \forall A.\ \Traces_A(P,\sigma_1)=\Traces_A(P,\sigma_2).
$
\end{sslhDefinition}

We first consider VellSCT's core fragment, without
declassifications or trusted external services.

\begin{sslhTheorem}[VellSCT soundness]
\label{thm:sct}
If VellSCT accepts $P$ in this fragment, the entry
contracts hold, and the allocation layouts and linking
environment satisfy the runtime assumptions, then
$P$ is SCT.
\end{sslhTheorem}

In other words, changing only Secret inputs does not
change the possible observations available to the same
adaptive attacker. The proof assumes valid allocation layouts and
unambiguous addresses for linked names. It matches
nondeterministic choices between executions to establish
equality of observable prefixes.

\paragraph{Proof sketch}
The proof maintains a relation between the two executions,
covering program counters, remaining computation trees,
SSA environments, heaps, and semantic bits. SSA values
satisfy the relation of Section~\ref{sec:contract-types}, while the heap
relation tracks memory domains, permissions, and public bytes.
The runtime invariant also matches observable faults.
A ready flag satisfies the coverage invariant for the common
semantic bit, while a pending edge records the guard
and successor until the corresponding update is executed.

We prove that each checked instruction preserves this
relation under Vellvm semantics. Public operands produce
matching branches and memory addresses. A load from normally
public memory may return different secret values after an
unsafe speculative access. Its Transient classification
prevents an unprotected observable use.
Protection makes related Transient values equal:
a clean flag preserves their normal equality, while a
set flag produces the same fixed result or matching failure.
Stores preserve the heap relation. At a join, all PHIs read the saved
predecessor environment before any destination is written. On a loop
backedge $\ell$, $x=\Phi[\ldots,\ell:y]$ and
$y=\Phi[\ldots,\ell:x]$ exchange the previous iteration's values;
sequential updates would incorrectly set both to the old $y$. The invariant
therefore relates the two snapshots and proves every selected incoming
operand related before constructing either successor environment. The flag
PHI additionally carries the token associated with that predecessor's
pending edge. This connects SSA value transport to the speculation bit,
rather than merely checking an integer PHI's shape.
Call and frame lemmas preserve the invariant across function boundaries;
definedness constraints match enabled operations and observable failures.

A weak simulation matches each observable event,
allowing silent allocation, environment, and concretization
steps. Composing these simulations gives matching
finite prefixes. Applying the argument in both directions
establishes trace-set equality.
The invariant, instruction cases, and trace composition are detailed in
Appendix~\ref{app:sct-invariant}--\ref{app:trace-composition}.

\paragraph{LLVM values and nondeterminism}
LLVM may leave value choices unresolved, and computations
may produce poison. Our proof follows these values through
evaluation and accounts for any observable failures
(Section~\ref{sec:llvm-challenges}). VellSCT rejects explicit
\texttt{undef} and \texttt{poison} operands, but these
restrictions do not prevent such values from arising during
execution. It permits \texttt{freeze i1} only immediately
before a conditional branch, with a Public operand. This
captures one defined Boolean for both the branch and its
MSF update.

Two executions may also make different choices without
leaking secrets. For example, allocation may choose different
public addresses even when the secret inputs are unchanged.
SCT allows these differences as long as changing the secrets
does not change which observations are possible. The backend
must preserve this independence when resolving under-defined
values at observable uses (Section~\ref{sec:limitations}). Architectural refinement alone is
insufficient: resolving a nondeterministic choice can shrink the two sets
of traces differently and break SCT. Security preservation therefore needs
secret-independent choice resolution in addition to functional refinement.

\subsection{Declassification and trusted external calls}
\label{sec:release-soundness}

Cryptographic programs sometimes intentionally disclose information, such as
whether authentication succeeded. This result may depend on Secret data,
so two executions with different secrets can legitimately reveal different
results at that point. A declassification policy $\mathcal R$ specifies
which values may be disclosed and where. We extend the security guarantee
to account for these declared disclosures.

We also support calls to a trusted external service, such as
\texttt{randombytes}, whose body lies outside the checked IR. The supported
interface takes a Public output-buffer address and length; its output bytes
are classified Secret. The declarations $\rho$ identify these calls, and
$S$ denotes the service implementation. We treat each call as an atomic
effect whose security contract is an explicit assumption.

\begin{sslhTheorem}[Declassification-aware VellSCT soundness]
\label{thm:release-sct}
Suppose VellSCT accepts $P$ under policy $\mathcal R$ and service
declarations $\rho$. Assume related initial states satisfying the entry
and global-memory contracts, valid layouts and linking, and that $S$
satisfies its relational service contract. For every finite execution
from either initial state, there is an execution from the other state
under the same attacker strategy such that:
\begin{enumerate}
\item both executions produce the same observations and reach related
configurations; or
\item their observations first differ at the same authorized
declassification site.
\end{enumerate}
\end{sslhTheorem}

Thus, matching observations are guaranteed up to the first differing
intentional disclosure. When the disclosed outcomes agree, the resulting
states remain related and the guarantee continues. Outcomes include
materialization failures as well as concrete values. Without
declassifications, the second case disappears and we recover SCT.

\paragraph{Proof idea}
We construct a matching execution under the same attacker strategy,
reusing Theorem~\ref{thm:sct}'s invariant and ordinary-instruction lemmas.
Public data and control agree; Secret values may differ.

Declassification needs a separate argument because it can make Secret
information observable. Its typing rule requires Public control, so the
matching execution reaches the same authorized site. If the disclosed
outcomes agree, the newly disclosed information agrees in both executions,
and the state relation is preserved. If they differ, this is the first
permitted difference because all earlier observations have already been
matched. During misspeculation, declassification reveals nothing and
retains the value's speculative classification, so matching continues
without authorizing a new disclosure.

At an external call, the buffer address and length are Public, so both
executions make the same request. The assumed service contract guarantees
that each response in the first execution has a matching response in the
second. On success, the resulting states remain related even though the
Secret output bytes may differ. On failure, both executions produce the
same observable fault.

Induction gives matching observations until an authorized difference;
exchanging initial states proves the reverse direction.
Appendix~\ref{app:declassification-proof} expands these proof cases, and
Appendix~\ref{app:formal-statements} gives the complete statement and premises.

\subsection{Architectural insertion refinement}

Hardening must prevent speculative leakage without
changing normal program behavior. During normal
execution, initialization produces a clean flag,
matching control-flow edges keep it clean, and
protection leaves values unchanged.

Let $\Harden_{\mathcal R,\rho}(P)=P'$ denote a successful
execution of the reference hardening entry point.
Let $E$ be an admissible context and observer, and let
$\Beh(P\oplus E,\sigma,O,\sigma',v)$ denote a complete
finite architectural execution, ending in a concrete observer
result or an observed fault. The optional result $v$ is absent
in the fault case.

\begin{sslhTheorem}[Architectural insertion preservation]
\label{thm:preservation}
Assume $\Harden_{\mathcal R,\rho}(P)=P'$ and that $S$
satisfies the architectural service contract.

For every admitted source entry, context, observer,
and initial world, every complete target behavior
$\Beh(P'\oplus E,\sigma,O,\sigma'_t,v)$ has either:
\begin{enumerate}
\item a source behavior
$\Beh(P\oplus E,\sigma,O,\sigma'_s,v)$
with the same observations and returned result,
and final-state refinement $\sigma'_s\preceq_v\sigma'_t$; or
\item a source execution that encounters undefined
behavior at a common observation prefix of $O$.
\end{enumerate}
\end{sslhTheorem}

Thus, when the source has no undefined behavior in the
admitted context, every target behavior is permitted by
the source. This result covers completed finite behaviors; it does not
establish divergence preservation. The phase ranking below rules out
unbounded bookkeeping within a simulated fragment, not every possible
whole-program divergence.

Here $\sigma'_s\preceq_v\sigma'_t$ preserves memory metadata
and requires every concrete typed-load outcome of the target
heap to be permitted by the source heap. Semantic bits agree;
on successful return, restored environments and frame stacks
also agree.
Symbolic expressions may differ, and target execution
may resolve nondeterministic choices of the source.
The architectural service contract uses this behavioral heap
relation; it is distinct from the relational security contract
in Theorem~\ref{thm:release-sct}.

Table~\ref{tab:root-premises} explains the initial-world and service premises;
Appendix~\ref{app:formal-details} states them in full.

\paragraph{Proof sketch}
The constructor first checks the source CFG and SSA
structure before generating fresh names.
The simulation relates each source instruction to its
generated sequence of operand protection, original
operations, and bookkeeping. The invariant tracks
original SSA values, generated Public values, the
clean flag, memory, suspended frames, and the current
phase of the generated sequence.
The proof uses LLVM's behavioral relations for
operands, including poison, undefined choices,
pointer provenance, and failures. A finite phase
ranking ensures progress through inserted instructions.
Separate invariants handle data and flag PHI nodes,
ensuring that each uses values from the correct
predecessor. The simulation extends across calls, returns, and
the observer. It also handles identity protection,
source undefined-behavior prefixes, pruning,
and the linked context.
Appendix~\ref{app:insertion-proof} gives the detailed fragment-simulation
argument.

\section{LLVM Implementation}
\label{sec:implementation}

\subsection{Frontend and extraction}

Clang transports scalar, initial-pointee and return confidentiality contracts
to LLVM. Programmers need not annotate Transient values or write flag
operations. Normalization inlines selected helpers, promotes stack variables,
lowers switches and supported memory intrinsics, and simplifies fixed layouts.
Matched comparisons use the same normalized input.

The OCaml driver parses LLVM into Vellvm's AST and invokes extracted
inference, insertion and VellSCT checking. The checker recomputes constraints
and validates the fixed point, operands, CFG/SSA structure, primitive
signatures and flag protocol. Emitted LLVM is verified, serialized, parsed
again and freshly checked.

An entry \emph{Public view} $v_P=\mathtt{protect}(v,m_0)$ gives a Public
replacement name dependent on initialization, retaining its pointee contract.
The original argument remains Transient under flow-insensitive typing.
For example, consumers use the new pointer name after the fence. The three
cores each have three such views. The backend can remove zero-mask arithmetic
while retaining an opaque ordering dependency until entry certification.

\subsection{Function calls and the MSF}

An ordinary call may overwrite the MSF register. Before calling, the caller
protects required Public arguments and executes \texttt{init\_msf} to stop
misspeculation. It initializes a fresh flag after return. The checker rejects
missing initialization and reuse of the old flag, so the calling convention
need not preserve an MSF register. Ordinary calls are restricted to non-exported helpers in the same module.
They must return \texttt{void}, contain no ordinary calls, and take at most
six Public integer or pointer arguments with normally Public pointee
contracts. Their stores require Public addresses and normally Public values.
Calls to trusted external services, such as \texttt{randombytes}, are
handled separately. The checker requires the output buffer's address and
length to be Public, so both executions make the same request, and relies
on the service contract of Section~\ref{sec:release-soundness} for the
service's behavior. Unlike ordinary calls, no extra initialization is
needed immediately before the call, because the service is assumed not
to leak secrets even when reached during misspeculation. After the call
returns, the caller initializes a fresh flag, since the old flag cannot
be relied on across the call. Each such initialization adds a barrier
cost to the call.

\subsection{Generating machine code}

\paragraph{Implementing the MSF operations}
\texttt{init\_msf} emits \texttt{LFENCE} followed by a move of zero on
x86-64. \texttt{LFENCE} stops later speculative execution until prior
instructions complete locally~\cite{intel-speculation-barriers}.
AArch64 uses \texttt{SB}, a speculation barrier, followed by zeroing.
Without \texttt{SB}, \texttt{DSB SY} completes earlier memory accesses and
\texttt{ISB} synchronizes instruction execution~\cite{arm-speculation-barriers}.

\texttt{set\_msf} computes a zero mask for a matching branch edge and
all ones for a mismatch, then ORs it with the old flag. The general x86
sequence copies the condition (\texttt{MOV}), subtracts or negates it
(\texttt{SUB}/\texttt{NEG}), then applies \texttt{OR}; AArch64
uses \texttt{SUB} and \texttt{ORR}. When existing x86 branch flags can be
reused safely, a conditional move (\texttt{CMOV}) instead selects the old
flag or all ones without recomputing the condition.

\texttt{protect} copies its input and ORs in the flag: zero preserves
the value, while all ones forces the fixed result. The instructions are
\texttt{MOV; OR} on x86 and \texttt{ORR} on AArch64. Default AArch64
lowering adds \texttt{CSDB} after updates and protection. This barrier restricts
use of predicted data without stopping branch speculation~\cite{arm-speculation-barriers}.
x86 updates and protection need no \texttt{LFENCE}.

\paragraph{Keeping the MSF in a register}
LLVM allocates a nonspillable virtual MSF register and propagates its tag
through copies. No physical register is permanently reserved; an attempted
spill rejects compilation. All 24 measured objects compile, but no general
failure rate is measured. LLVM SLH already uses a virtual predicate; we add
mandatory spill rejection.

\paragraph{Handling ordinary stack spills}
Ordinary spills introduce reloads after IR checking. A Public stack address
alone does not certify their contents. The x86 backend certifies supported
Public store/reload paths; otherwise it masks Transient reloads feeding Public
uses with the current MSF, rejecting unsupported cases. AArch64 uses a barrier
fallback or optional GPR masks. Appendix~\ref{app:backend-proof} gives the
store-width, initialization and aliasing conditions, including overlapping writes.

\paragraph{Security after code generation}
\label{lem:backend-transport}
The Rocq proofs stop at LLVM IR; MSF/reload checks do not discharge all backend
obligations. Later passes must preserve protection dependencies and resolve
under-defined values independently of secrets at observable uses, under the
compiler and hardware conditions in Appendix~\ref{app:backend-proof}.
The evaluated pipeline disables select optimization and x86 conditional-move
conversion to avoid new branches.

\section{Evaluation}
\label{sec:evaluation}

\paragraph{Experimental setup.}
We evaluate 24 cryptographic operations from TweetNaCl, libsodium, HACL*,
and Kyber. They include MACs, stream ciphers, Secretbox, SHA/SHAKE,
X25519, Kyber512/768, and Salsa/ChaCha-family cores. For each operation,
we compare four versions: unprotected code, LLVM SLH, Ultimate SLH, and
VellSSLH. Each comparison uses the same cryptographic implementation,
inputs, and test driver. Appendix~\ref{app:benchmark-preparation} gives
the source-preparation and compilation details.

We measure native runtime on a bare-metal Linux machine with an
Intel Core Ultra~5~335, using the same performance core (CPU~2).
Unprotected code, LLVM SLH, and VellSSLH use our LLVM~23.0.0git build.
We compare these three versions in nine trials, varying their code layout
and measurement order in each trial to reduce sensitivity to layout and
frequency changes. Ultimate SLH uses LLVM~16 and is measured in a separate
set of nine trials on the same core.

Before timing, we warm up the code and run functional checks. We time
repeated function calls, including the common test loop and output
checksum; calibration, reference computations, and full-output checks
are excluded. We take the median time in each trial, then report
the median across the nine trials. For unprotected code and LLVM SLH,
percentage changes use ratios from matching trials, combined with a
geometric mean. For Ultimate SLH, they use the ratio of the two reported
medians. The 95\% confidence intervals use whole trials as samples, and
we retain every trial.

\subsection{RQ1: Applicability to Cryptographic C}
\label{sec:eval-applicability}

All 24 hardened modules pass VellSCT and the functional output checks.
Table~\ref{tab:crypto-summary} reports results for 16\,KiB messages where
the operation accepts variable-length input. X25519 uses 32-byte scalars,
the fixed cores use 16-byte inputs, and KEM measurements cover complete
operations. SHAKE produces 64 output bytes. The artifact includes tested
message sizes from 16 bytes to 64\,KiB. Whole-library support remains
limited by source preparation, unsupported instructions, dispatch code,
and Secret select guards.

\paragraph{Analysis cost.}
Table~\ref{tab:analysis-cost} measures analysis time and peak memory on an
Apple M4 with 16\,GiB RAM. Each entry uses a fresh process and includes
parsing and LLVM validation. \emph{Check} runs VellSCT on hardened IR;
\emph{Insert} includes inference and protection removal, starting from
prepared IR. The table gives its size in KiB and peak resident memory (RSS).

The table uses a 180-second limit. The runs that produced the binaries
allowed 1,200 seconds for Kyber and 1,800 for the broader suite. X25519 took
409 seconds for insertion and 127 seconds for rechecking; Kyber512
decapsulation took 769 and 339 seconds. These modules passed with larger
limits despite the table's timeouts. Inlining enlarges the IR, over which
the checker repeatedly processes constraints. We have not timed this
separately from parsing and validation.

\begin{table}[!tbp]
\centering\small
\setlength{\tabcolsep}{5pt}
\begin{tabular}{lrrrrr}
\toprule
Input & IR (KiB) & Check (s) & RSS (MiB) & Insert (s) & RSS (MiB) \\
\midrule
Salsa20 core & 23.5 & 0.55 & 20.0 & 3.95 & 20.0 \\
HSalsa20 core & 16.8 & 0.22 & 19.9 & 1.56 & 19.9 \\
HChaCha20 core & 16.8 & 0.40 & 19.9 & 4.56 & 20.0 \\
Poly1305 & 58.5 & 4.55 & 20.4 & 26.87 & 20.3 \\
SHA3-256 & 40.2 & 0.91 & 20.2 & 6.58 & 20.2 \\
X25519 scalar & 160.4 & 70.65 & 145.8 & $\geq180$ & 153.2 \\
Kyber512 decapsulation & 343.9 & $\geq180$ & 31.1 & $\geq180$ & 122.3 \\
\bottomrule
\end{tabular}
\caption{Analysis time and peak memory on an Apple M4 (16 GiB); $\geq180$ denotes a timeout.}
\label{tab:analysis-cost}
\end{table}

\paragraph{Annotation effort.}
Programmers specify Public and Secret contracts. VellSSLH infers
Transient types and inserts the MSF operations. The Poly1305 wrapper uses
five annotations; SHA3-256 and X25519 scalar multiplication each use seven,
including unused test-interface parameters. Kyber512
decapsulation uses four annotations and two sampler declassification
expressions, which become four sites after inlining. We did not measure
programmer time.

Our Kyber implementation is Round~3 v3.0, rather than FIPS~203 ML-KEM.
Its policy makes matrix-sampler values public while keeping noise
seeds, keys, and shared secrets Secret. The conservative memory analysis
may require some releases; we have not measured which separate tracking
of each allocation would avoid.
Secretbox open also declassifies whether authentication succeeded.
Security depends on these declared policies and the external-service
contracts. Key generation and encapsulation use the
operating-system random-number generator.

\subsection{RQ2: Performance Compared with Existing SLH}
\label{sec:eval-performance}

Table~\ref{tab:crypto-summary} compares unprotected code, LLVM SLH,
full Ultimate SLH, and VellSSLH on the same host and core.

\begin{table}[!tbp]
\centering\small
\setlength{\tabcolsep}{2.5pt}
\begin{tabular}{lrrrrrrrc}
\toprule
Operation & Bare & \shortstack{LLVM\\SLH} & \shortstack{Ultimate\\SLH} & VellSSLH & \shortstack{$\Delta$ vs\\bare (\%)} & \shortstack{$\Delta$ vs\\SLH (\%)} & \shortstack{$\Delta$ vs\\Ultimate (\%)} & Spec. \\
\midrule
Poly1305 & 7.429 & 8.420 & 8.452 & 7.775 & +4.93 & -7.49 & -8.01 & $\checkmark$ \\
SipHash-2-4 & 5.369 & 5.439 & 5.422 & 5.373 & +0.08 & -1.18 & -0.91 & $\checkmark$ \\
SipHash128 & 5.381 & 5.451 & 5.429 & 5.325 & -1.04 & -2.27 & -1.91 & $\checkmark$ \\
ChaCha20 & 24.429 & 23.896 & 24.206 & 23.543 & -3.52 & -1.11 & -2.74 & $\checkmark$ \\
Salsa20 & 38.852 & 52.146 & 67.620 & 45.689 & +17.95 & -12.24 & -32.43 & B \\
XSalsa20 & 39.812 & 52.605 & 72.440 & 44.612 & +12.58 & -15.01 & -38.42 & B \\
Secretbox encrypt & 341.196 & 402.309 & 819.762 & 362.844 & +6.34 & -9.81 & -55.74 & M \\
Secretbox open & 159.631 & 240.762 & 535.922 & 223.031 & +40.79 & -7.56 & -58.38 & $\checkmark$ \\
SHA3-256 & 25.670 & 26.182 & 27.707 & 26.497 & +3.32 & +1.22 & -4.37 & M \\
SHA3-512 & 47.900 & 48.690 & 50.492 & 49.184 & +2.68 & +1.11 & -2.59 & M \\
SHA-512 & 31.289 & 36.477 & 42.784 & 33.644 & +7.63 & -7.78 & -21.36 & M \\
SHAKE128 & 21.071 & 21.715 & 22.982 & 21.750 & +3.49 & +0.19 & -5.36 & M \\
SHAKE256 & 25.735 & 26.530 & 28.291 & 26.664 & +3.82 & +0.67 & -5.75 & M \\
X25519 scalar & 589.145 & 776.418 & 844.987 & 600.199 & +1.86 & -22.69 & -28.97 & M \\
X25519 base & 589.276 & 776.151 & 844.236 & 601.333 & +2.03 & -22.56 & -28.77 & M \\
Kyber512 keygen & 15.791 & 18.360 & 23.531 & 18.869 & +19.71 & +2.83 & -19.81 & M \\
Kyber512 encap. & 18.881 & 22.616 & 28.501 & 22.351 & +18.28 & -1.24 & -21.58 & M \\
Kyber512 decap. & 21.005 & 25.318 & 33.572 & 24.382 & +16.22 & -3.53 & -27.38 & M \\
Kyber768 keygen & 26.253 & 31.199 & 40.697 & 31.159 & +18.59 & -0.01 & -23.44 & M \\
Kyber768 encap. & 29.476 & 35.195 & 46.375 & 35.868 & +21.57 & +1.80 & -22.66 & M \\
Kyber768 decap. & 32.437 & 39.501 & 52.823 & 39.374 & +21.05 & -0.44 & -25.46 & M \\
Salsa20 core & 0.081 & 0.083 & 0.084 & 0.087 & +6.86 & +3.83 & +3.25 & $\checkmark$ \\
HSalsa20 core & 0.081 & 0.077 & 0.078 & 0.084 & +3.24 & +8.02 & +7.73 & $\checkmark$ \\
HChaCha20 core & 0.083 & 0.082 & 0.083 & 0.088 & +6.38 & +6.95 & +6.09 & $\checkmark$ \\
\bottomrule
\end{tabular}
\caption{Representative bare-metal runtime ($\mu$s/call). Negative changes mean faster. Bounded checks: $\checkmark$ passed; M model-limited; B budget-limited.}
\label{tab:crypto-summary}
\end{table}

\paragraph{Comparison with LLVM SLH}
VellSSLH is 4.03\% faster across the 24 operations, using the geometric
mean of runtime ratios. Excluding the two X25519 operations gives a
2.13\% improvement. The measured ratios favor VellSSLH in 15 operations.
The 95\% confidence intervals support an improvement in 13 and a slowdown
in eight. Three comparisons remain inconclusive: SHAKE128 and Kyber768
key generation and decapsulation. These intervals describe variation on
this machine; we have not repeated the study on other processors.

X25519 scalar multiplication and base multiplication are 22.69\% and
22.56\% faster than LLVM SLH, with 1.86\% and 2.03\% overhead over
unprotected code. Salsa20 and XSalsa20 are 12.24\% and 15.01\% faster,
and SHA-512 is 7.78\% faster. The three fixed cores are 3.83--8.02\%
slower. Secretbox open is 7.56\% faster than LLVM SLH but still 40.79\%
slower than unprotected code. Some measurements, including ChaCha20,
are faster than the unprotected version; code layout can affect these
timings, so this does not show that hardening has no cost.
Section~\ref{sec:eval-backend} reports a separate experiment in which
backend settings also improve SHA3-256 and Kyber768 encapsulation.

\paragraph{Comparison with Ultimate SLH}
We build Ultimate SLH~\cite{ultimate-slh} with LLVM~16 and all published
protection options enabled, using equivalent prepared IR and the same
test drivers. VellSSLH is faster in 21 operations, with a 19.65\% lower
geometric-mean runtime. All 21 confidence intervals support an improvement.
The three fixed cores remain slower.

Ultimate SLH also protects instructions whose execution time can depend
on their operands. Our checker requires division operands to be Public:
a Transient divisor needs protection, and a Secret divisor is rejected.
This addresses part of that concern, but our security guarantee does not
cover all variable-time instructions. To check whether Ultimate's extra
protection affects this comparison, we compile all 24 routines with that
option disabled. Every resulting object is byte-for-byte identical to
its fully protected version. The timings use the fully protected versions.

The tools use different LLVM versions and available options, and their
timings come from separate runs. The difference in runtime therefore
cannot be attributed entirely to the hardening algorithm.
Appendix~\ref{app:ultimate-comparison} gives the confidence intervals;
the artifact includes measurements at additional input sizes. We have
not validated the Ultimate binaries with Spectector.

\subsection{RQ3: Benefit of Selective Protection}
\label{sec:eval-selectivity}

To measure the benefit of selectivity, we compare two ways of inserting
protection within VellSSLH. Conservative placement protects potentially
sensitive uses based on the instruction form. Selective placement uses
inferred types to remove unnecessary protections. Both use the same
prepared IR, MSF operations, one entry fence, nonspillable flag, and
backend checks. We disable entry-view reuse for this experiment; the main
performance table uses the default configuration.

For the Salsa20, HSalsa20, and HChaCha20 cores, selective placement reduces
the number of IR protection calls from 149, 85, and 85 to three per core.
The function sizes fall from 2,387, 1,388, and 1,299 bytes to 1,268, 874,
and 808 bytes. Table~\ref{tab:controlled-placement} reports runtime.
\emph{Bare} is the unprotected version; \emph{Same flags} is unprotected
code compiled with VellSSLH's backend restrictions.

\begin{table}[!tbp]
\centering\small
\setlength{\tabcolsep}{4pt}
\begin{tabular}{lrrrrrr}
\toprule
Core & Bare & Same flags & LLVM SLH & Consv. & Selective & Ratio [95\% interval] \\
\midrule
Salsa20 & 0.081 & 0.081 & 0.084 & 0.095 & 0.090 & 0.947 [0.943, 0.950] \\
HSalsa20 & 0.081 & 0.081 & 0.077 & 0.090 & 0.086 & 0.954 [0.951, 0.957] \\
HChaCha20 & 0.083 & 0.083 & 0.082 & 0.091 & 0.089 & 0.977 [0.970, 0.983] \\
\bottomrule
\end{tabular}
\caption{Protection-placement comparison ($\mu$s/call); ratios compare Selective with Conservative placement.}
\label{tab:controlled-placement}
\end{table}

After functional checks, we run nine trials with randomized layouts.
The 95\% intervals use 10,000 resamples of whole trials, with none
discarded. Selective placement is faster by 5.34\%, 4.59\%, and 2.26\%,
respectively, and wins in all nine trials. The two unprotected controls
have identical function bytes, so their timing differences show the
effect of layout and measurement variation.

With these settings, selective VellSSLH remains 7.47\%, 10.22\%, and
7.61\% slower than LLVM SLH. Selectivity reduces code size and runtime,
but both VellSSLH versions still pay for the entry barrier. These cores
take roughly 80\,ns, so that fixed cost is a large part of each call.
We have not repeated this experiment on X25519, full Salsa20, or SHA-512;
their improvements over LLVM SLH may also come from other differences
between the compilers.

\subsection{RQ4: Backend Costs and Configuration}
\label{sec:eval-backend}

\paragraph{Protection and flag costs.}
Table~\ref{tab:protection-counts} counts three kinds of VellSSLH operations:
protection calls in the checked IR, protection added for stack reloads,
and initialization fences. The LLVM SLH column counts the register
results it hardens after loads. We obtain those counts by replaying its
pass and checking that it reproduces the measured assembly, so ordinary
cryptographic AND/OR instructions are excluded. The LLVM SLH objects
contain no \texttt{LFENCE}. These counts refer to places in the code;
a protection inside a loop can execute many times.

The counts come from different compilation stages: VellSSLH's IR calls
may still be optimized during code generation. Its memory analysis also
uses conservative labels that do not change at each program point.
This can cause additional protections in Secretbox open and SHA3/SHAKE.
SHA3/SHAKE are slower than LLVM SLH, while Secretbox open remains faster
despite its high overhead over unprotected code. Tracking changes in
memory labels more precisely could reduce unnecessary protection.

\begin{table}[!tbp]
\centering\small
\caption{Static protection and initialization sites for the default objects.}
\label{tab:protection-counts}
\begin{tabular}{lrrrr}
\toprule
Operation & \shortstack{LLVM SLH\\post-load regs} & \shortstack{VellSSLH\\IR protects} & \shortstack{VellSSLH\\late reloads} & \shortstack{VellSSLH\\LFENCEs} \\
\midrule
Poly1305 & 63 & 4 & 3 & 1 \\
SipHash-2-4 & 10 & 4 & 0 & 1 \\
SipHash128 & 10 & 4 & 0 & 1 \\
ChaCha20 & 28 & 13 & 0 & 1 \\
Salsa20 & 32 & 6 & 6 & 1 \\
XSalsa20 & 32 & 6 & 4 & 1 \\
Secretbox encrypt & 76 & 5 & 36 & 1 \\
Secretbox open & 74 & 147 & 57 & 1 \\
SHA3-256 & 6 & 39 & 6 & 1 \\
SHA3-512 & 6 & 71 & 6 & 1 \\
SHA-512 & 25 & 25 & 1 & 1 \\
SHAKE128 & 10 & 13 & 12 & 1 \\
SHAKE256 & 10 & 13 & 12 & 1 \\
X25519 scalar & 129 & 11 & 3 & 1 \\
X25519 base & 129 & 9 & 3 & 1 \\
Kyber512 keygen & 119 & 87 & 50 & 3 \\
Kyber512 encap. & 211 & 51 & 62 & 2 \\
Kyber512 decap. & 306 & 88 & 53 & 1 \\
Kyber768 keygen & 111 & 89 & 49 & 3 \\
Kyber768 encap. & 200 & 52 & 70 & 2 \\
Kyber768 decap. & 299 & 91 & 65 & 1 \\
Salsa20 core & 12 & 3 & 0 & 1 \\
HSalsa20 core & 12 & 3 & 0 & 1 \\
HChaCha20 core & 12 & 3 & 0 & 1 \\
\bottomrule
\end{tabular}
\end{table}

Small messages do little computation, so fixed costs make up a larger
fraction of runtime. An added cost $\delta$ gives
$(100\delta)/T_{\mathrm{bare}}(n)$ percent overhead at message length $n$.
VellSSLH executes an entry \texttt{LFENCE} even when it needs few masks;
LLVM SLH instead carries its poison state through the stack pointer.
Register pressure, register saves, and protection of stack reloads can
also reduce the benefit of selective insertion.

X25519 scalar/base use 11/9 IR protections and three added reload
protections, compared with 129 hardened load results in LLVM SLH.
Salsa20/XSalsa20 use six IR protections and six/four reload protections,
compared with 32; SHA-512 uses 25 and one, compared with 25. Each of these
VellSSLH routines uses one fence, while the LLVM SLH versions use none.
The counts show how the generated code differs, but we have no hardware
counter measurements that isolate the causes of the X25519 speedup.

\paragraph{Backend configuration experiment.}
Table~\ref{tab:backend-tuning} tests two compiler options on four hash and
three KEM operations, using nine trials with randomized layouts. The IR,
checker, insertion algorithm, compiler sources, and proofs stay the same.
One option separates loads from the instructions using them, allowing
the backend to check whether spilled Public values remain Public when
reloaded. This can reduce masks but add instructions. The other releases
a scratch register for other uses.

We rerun the unprotected and LLVM SLH controls in each trial; their times
can differ from the main table, as for Kyber768 key generation. The
initialization and call protocols remain unchanged, and all backend and
functional checks pass. Appendix~\ref{app:backend-tuning-controls} shows
that releasing the register helps SHA3-256; separating loads alone gives
no clear improvement.

\begin{table}[!tbp]
\centering\small
\setlength{\tabcolsep}{2.5pt}
\begin{tabular}{lrrrrrr}
\toprule
Operation & \shortstack{LLVM\\SLH} & \shortstack{VellSSLH\\default} & \shortstack{VellSSLH\\tuned} & \shortstack{$\Delta$ vs default (\%)\\95\% interval} & \shortstack{$\Delta$ vs SLH (\%)\\95\% interval} & \shortstack{Late\\masks} \\
\midrule
SHA3-256 & 26.216 & 26.435 & 25.967 & $-1.76\;[-2.15,-1.37]$ & $-0.73\;[-1.03,-0.42]$ & $6\to0$ \\
SHA3-512 & 48.573 & 49.323 & 48.654 & $-1.10\;[-1.45,-0.67]$ & $+0.38\;[+0.12,+0.64]$ & $6\to0$ \\
SHAKE128 & 21.745 & 21.793 & 21.831 & $+0.04\;[-0.47,+0.48]$ & $+0.36\;[-0.08,+0.77]$ & $12\to3$ \\
SHAKE256 & 26.446 & 26.802 & 26.757 & $-0.45\;[-0.82,-0.06]$ & $+0.96\;[+0.66,+1.24]$ & $12\to3$ \\
Kyber512 keygen & 18.439 & 18.888 & 18.784 & $-0.55\;[-0.81,-0.30]$ & $+1.98\;[+1.62,+2.35]$ & $50\to22$ \\
Kyber768 keygen & 31.177 & 31.177 & 31.147 & $-0.14\;[-0.24,-0.06]$ & $0.00\;[-0.13,+0.12]$ & $49\to24$ \\
Kyber768 encap. & 35.429 & 36.008 & 34.832 & $-3.19\;[-3.40,-2.99]$ & $-1.69\;[-2.01,-1.34]$ & $70\to30$ \\
\bottomrule
\end{tabular}
\caption{Backend tuning: runtime ($\mu$s/call), paired changes with 95\% intervals, and late reload masks.}
\label{tab:backend-tuning}
\end{table}

Six operations improve over default VellSSLH. SHA3-256 and Kyber768
encapsulation become 0.73\% and 1.69\% faster than LLVM SLH; SHAKE128
and Kyber768 key generation remain inconclusive. SHA3-512, SHAKE256, and
Kyber512 key generation remain slower. SHA3-512 is still 0.38\% slower
despite having no added reload masks. We have not tested every option
combination or checked these changed binaries with Spectector; RQ5
covers the default binaries.

\subsection{RQ5: Functional and Binary Security Validation}
\label{sec:eval-security}

We test the generated binaries for functional correctness and use
Spectector to check security within a fixed exploration bound.

\paragraph{Functional correctness.}
Before timing, we compare outputs with reference implementations or
other versions. Tests cover known vectors, boundary inputs, alignment,
buffer guard bytes, input preservation, and supported overlapping buffers.
Each trial runs 59,436 checks across 72 default objects: 534,924 passing
cases across 648 linked binaries. Kyber tests check KEM consistency and
boundaries, without independent known-answer validation. Object and
function hashes link tests, timings, and binary analysis to the same code.

\paragraph{Bounded binary analysis.}
We use Spectector~\cite{spectector} with our x86 translation adapter and
separately tested extensions to its instruction model
(Appendix~\ref{app:binary-coverage}). Each query allows a speculation
window of 64, 20,000 execution steps, and 20 seconds. Selected retries
allow up to 50,000 steps and 15 minutes. Eight VellSSLH binaries pass
all selected queries: Poly1305, SipHash-2-4, SipHash128, ChaCha20,
Secretbox open, and the Salsa20, HSalsa20, and HChaCha20 fixed cores.
The three LLVM SLH fixed cores also pass. These results apply to the
tested binaries and configurations, not to every input or execution.

The remaining 16 binaries have no verdict. Full Salsa20 and XSalsa20
exceed the resource limits. The other 14 contain instructions that the
adapter does not fully model, including XMM operations, word-width
operands, rotations in memory, and carry/borrow operations. LLVM can
generate XMM instructions even from scalar IR.
Appendix~\ref{app:binary-coverage} lists the obstacle for each binary.
These are limits of the binary analysis, not VellSCT rejections or
detected leaks. The binary checks also do not establish the validity of
declassification policies or external-service contracts.

\section{Related Work}
\label{sec:related-work}

\paragraph{Speculative security and selective hardening}
Cauligi et al.~\cite{sct-foundations} formulate SCT for directed speculative
and out-of-order execution in Pitchfork.
Barthe et al.~\cite{high-assurance-spectre} combine speculative memory safety
and constant-time reasoning in Jasmin. Our LLVM model compares possible
prefixes to accommodate nondeterminism.

\emph{Spectre Declassified}~\cite{spectre-declassified} introduces selective
SLH and explains how intentional disclosure interacts with speculation.
\emph{Typing High-Speed Cryptography against Spectre v1}~\cite{jasmin-sslh}
provides normal/speculative labels, MSF typestate and three hardening
primitives. VellSSLH addresses C-derived SSA, predecessor-sensitive flag PHIs
and LLVM's symbolic values, poison, provenance and faults.

Patrignani and Guarnieri~\cite{exorcising} develop secure-compilation criteria
for Spectre v1 mitigations and analyze strong SLH. Ultimate SLH~\cite{ultimate-slh}
also protects variable-time arithmetic. FSLH~\cite{fslh} generalizes selective
and ultimate hardening with mechanized relative-security guarantees and SCT
corollaries. We prove checker soundness and architectural refinement for the
LLVM/Vellvm fragment, observing branches, addresses and failures.

\paragraph{Automatic placement and LLVM implementations}
Blade~\cite{blade} uses transient/stable types and minimum dataflow cuts to
synthesize protections in Cranelift/WebAssembly. VellSSLH uses inferred
labels, reusable Public views and checked pruning, with an independent
checker validating placement. LightSLH~\cite{lightslh} uses abstract
interpretation and a hardening-aware fixed point; its compiler-inserted-load
examples motivate our late reload treatment. Serberus/LLSCT~\cite{serberus}
protects constant-time crypto against broader speculation primitives under
hardware control-flow assumptions. LLVM SLH~\cite{llvm-slh} supplies our
production performance baseline. Broader comparisons with LightSLH under matched LLVM settings and with
optimized libjade on shared primitives remain future work.

\paragraph{Preserving security through compilation}
CompCertCT~\cite{compcertct} proves constant-time preservation using
leakage-preserving, erasing and transforming simulations. Jasmin's original
compiler paper~\cite{jasmin-2017} gives a manual pass-wise constant-time
argument. For speculation, Arranz Olmos et al.~\cite{sct-preserving-compilation}
exhibit spill/memory-reuse counterexamples and mechanize preservation for
nine passes of a Jasmin-inspired compiler. SNIP~\cite{snip} develops
speculative non-interference preservation with directive-adapting simulations,
including a register-allocation counterexample and repair.
Hardware--software contracts~\cite{speculation-contracts} provide a framework
for stating the barrier and dependency assumptions we require.
Jasmin's Spectre-RSB protection~\cite{jasmin-rsb} treats return prediction and
broader variants; our post-call fences do not constitute an RSB security claim. Our backend
schema identifies the corresponding LLVM flag, reload, frame, rewrite and
choice-selection obligations.

\paragraph{LLVM semantics and independent checking}
The original Vellvm development and Zhao's thesis establish SSA and verified
transformation foundations~\cite{vellvm-popl,zhao-thesis}. Modern Vellvm uses
interaction trees and modular handlers for effects, under-defined values and
nondeterminism~\cite{vellvm-modular,vellvm-experience}. We add speculative
directives and observations while preserving those value/memory interfaces.
Spectector~\cite{spectector} independently analyzes bounded speculative
information flow in assembly. Its instruction model and exploration bounds
determine which of our objects it can decide.

\section{Limitations and Trust Boundaries}
\label{sec:limitations}
The scalar fragment excludes SIMD, inline assembly, pointer-valued memory,
atomics, volatile effects and unrestricted calls. Conservative memory and
select rules limit acceptance; resource exhaustion yields unknown.
Contracts, declassifications and external-service specifications are trusted
policies, not proofs of cryptographic correctness or disclosure safety.
Annotation transport, normalization, extraction, parsing and printing are
also trusted. Backend checks cover specific MSF/reload conditions, not all
LLVM rewrites, stack/ABI behavior, directive adaptation or hardware contracts.
In particular, resolving \texttt{IMPLICIT\_DEF} from stale Secret register
contents could leak despite passing the MSF checks.

The forward sticky-bit model excludes rollback, indirect-branch attacks,
store bypass and other timing channels. Bounded binary analysis establishes
no verdict for unsupported or exhausted queries. Performance covers selected
portable implementations on one Intel host with randomized layouts, not
whole libraries, optimized production crypto or other microarchitectures.
AArch64 lowering is implemented but unmeasured. Other LLVM frontends,
including Rust, could reuse the analysis after transporting secrecy contracts;
panics, aggregates and overflow intrinsics may require new rules and proofs.

\paragraph{AI Use}
We used OpenAI's Codex and Anthropic’s Claude for implementation assistance, proof repair,
experiment automation, artifact preparation, consistency checks, data
analysis, and drafting and editing prose. Formal guarantees rely on
Rocq-checked proofs. We take full responsibility for the correctness and content of this work.

\section*{Data-Availability Statement}
The research artifact will be made publicly available after acceptance of
this paper. It will include the Rocq proof development and dependencies,
compiler modifications, extraction scripts, benchmark harnesses and objects,
raw measurements, and validation reports, together with build and
reproduction instructions.

\bibliographystyle{ACM-Reference-Format}
\bibliography{references}

\clearpage
\appendix
\section{IR Proof Structure}
\label{app:formal-details}

We first state the IR theorems and their assumptions, then describe the
state relations and instruction lemmas used to prove them.

\subsection{Formal statements and assumptions}
\label{app:formal-statements}

The three main theorems use compatible implementations of Vellvm's LLVM
parameters, memory, bytes, concretization, and shared runtime, together
with the CFG and source interfaces. The statements below include their
outer assumptions and use Vellvm's call semantics.

\begin{table}[!htbp]
\centering\small
\begin{tabular}{lp{0.73\linewidth}}
\toprule
Premise & Meaning \\
\midrule
Runtime & Addresses, provenance, layouts, bytes, and concretization satisfy the Vellvm interfaces. \\
Check/Prepare & The executable checker accepts, and the entry contracts and input kinds are decoded successfully. \\
Init & Execution starts normally in related states, with readable arguments, admissible bytes, and agreement on declared public memory. \\
Layout & Every allocated LLVM type has a valid nonvoid layout. \\
Link & Function addresses identify names unambiguously and agree with global lookup. \\
GlobalView & Tracked globals are bound to addresses and satisfy their public-memory contracts. \\
Safe/Arch & The external service satisfies its security or architectural-refinement contract, respectively. \\
World & The entry, observer, calling convention, and global addresses satisfy the architectural execution requirements. \\
\bottomrule
\end{tabular}
\caption{Assumptions of the main theorems.}
\label{tab:root-premises}
\end{table}

\paragraph{Executions and initial indistinguishability}
Write $C\execprefix{A,h}{O}C'$ for an execution from $C$ to $C'$ with
finite trace $O=o_1\cdots o_n$. The silent steps preceding $o_i$ and the
step producing it use directive $A(h\cdot o_1\cdots o_{i-1})$, chosen
from the observation history. The possible traces from $C$ are
\[
 \Traces_A(C)=\{O\mid\exists C'.\ C\execprefix{A,\epsilon}{O}C'\}.
\]
The relation $\mathsf{Init}(I,V,\sigma_1,\sigma_2,\vec a)$ requires both
states to start outside speculation with empty local environments and
frame stacks. Their heaps have matching metadata, agree on store
identifiers and bytes at visible addresses $V(p)$, and match byte reads
in both directions. All readable heap bytes must be admissible.

The pairs in $\vec a$ list the prepared inputs in order. Each pair
satisfies $\approx^0_\tau$, and both values are readable at the expected
kind. For a normally Public memory input, $V$ includes every allocated
byte reachable through any concretization of the pointer's provenance
in either heap.

A readable scalar is correctly typed and yields an integer under
nondeferred concretization. A readable pointer is an address or a
correctly typed deferred value. An admissible byte comes from an in-bounds
position of a concrete integer, an address, or a correctly typed deferred
parent value.

The premise $\mathsf{Layout}(P)$ checks recursive nonvoid validity of every
allocated LLVM type in $P$. Let $T$ be the function-address table and $g$ the global
environment. $\mathsf{Link}(T,g)$ requires
$(n,p),(n',p)\in T\Rightarrow n=n'$ and
$g(n)=\mathtt{Addr}(p)\Rightarrow(n,p)\in T$.
The configuration $\mathsf{Run}_{P,T,g}(f,\vec v,\sigma)$ executes the
normalized meaning of $f$ with the module's linked callees. The prepared
return contract determines which returned values the observer can see.

\paragraph{Soundness of accepted IR}
For every $P,f,m,r,I,T,g,V,\sigma_1,\sigma_2,\vec a$,
\[
\begin{gathered}
 \mathsf{Check}(P)=\mathsf{true}\ \land\ f\in\mathsf{Defs}(P)
 \ \land\ \mathsf{Layout}(P)\ \land\ \mathsf{Link}(T,g)\\
 {}\land\ \mathsf{Mode}(m,P,f)=\mathsf{Some}(r)
 \ \land\ \mathsf{Prepare}(m,P,f)=\mathsf{Some}(I)
 \ \land\ \mathsf{Init}(I,V,\sigma_1,\sigma_2,\vec a)\\
 \Longrightarrow\quad\forall A.\
 \Traces_A(\mathsf{Run}(f,\pi_1\vec a,\sigma_1))
 =\Traces_A(\mathsf{Run}(f,\pi_2\vec a,\sigma_2)).
\end{gathered}
\]
Mode checking and preparation read the module's attribute groups.
The projection $\pi_j$ selects the $j$th value from each argument pair.
The premises require acceptance and valid initial conditions; the theorem
proves that the function bodies have matching observable behavior.

\paragraph{Declassification-aware soundness}
For policy $\mathcal R$, designated-service names $\rho$, and service $s$,
$\mathsf{Safe}(s)$ is the atomic security contract. The service receives
only the heap and request. On success, it writes only the writable,
consecutive output range and preserves other bytes, metadata, and saved
caller frames. Equal public requests in related states have matching
responses: success with related states and Secret output bytes, or
the same failure.
For every $s,\mathcal R,\rho,P,D,f,I,T,g,V,\sigma_1,\sigma_2,\vec a$,
\[
\begin{gathered}
 \mathsf{Safe}(s)\land\mathsf{Check}_{\mathcal R,\rho}(P)=\mathsf{true}
 \land\mathsf{Readonly}(P)=\mathsf{Some}(D)\land f\in\mathsf{Defs}(P)\\
 {}\land\mathsf{PrepareExt}(P,f)=\mathsf{Some}(I)
 \land\mathsf{Layout}(P)\land\mathsf{Link}(T,g)
 \land\mathsf{Init}(I,V,\sigma_1,\sigma_2,\vec a)\\
 {}\land\mathsf{GlobalView}(D,V,g,\sigma_1)
 \land\mathsf{GlobalView}(D,V,g,\sigma_2)
 \quad\Longrightarrow\quad \mathsf{Coupled}(C_1,C_2).
\end{gathered}
\]
$C_j$ uses the declassification-aware semantics on $\pi_j\vec a$,
with the module's functions and designated service linked in.
$\mathsf{GlobalView}$ requires address bindings for tracked globals;
the provenance-accessible bytes of normally Public memory must lie in
the visible region. $\mathsf{Coupled}$ supplies a relation containing
$(C_1,C_2)$ that matches silent and ordinary steps in both directions,
preserves equal semantic bits, and matches declassification sites.
At related sites, \emph{every pair of actual steps with equal
declassification outcomes} must leave related states.

The trace corollary holds in both directions. For all $A,h,O,C'_1$,
\[
 C_1\execprefix{A,h}{O}C'_1\ \Longrightarrow\
 \bigl(\exists C'_2.\ C_2\execprefix{A,h}{O}C'_2
       \land\mathsf{Coupled}(C'_1,C'_2)\bigr)
 \ \lor\ \mathsf{Diff}_{A,h}(C_1,C_2,O).
\]
In $\mathsf{Diff}$, the first difference is an authorized release:
$\mathsf{declass}(\ell,u)$ and $\mathsf{declass}(\ell,u')$, with $u\ne u'$.
The left trace includes this release and any later observations.
Earlier observations must agree. A release outcome is a concrete value
or a materialization failure.

\paragraph{Architectural behavior and insertion refinement}
$\Beh_{s,\mathcal R,\rho,T,g}(P,K,f,\vec v,o,\vec e,\sigma,O,\sigma',z)$
describes a completed finite execution using only $\Step$. It executes
$f$ linked with $P{+}K$, passes any nonvoid return value and extra arguments
$\vec e$ to observer $o$, then materializes the observer's result.
The optional result is $z$. Observing poison produces undefined behavior;
out-of-memory and other faults remain observable.

The final-state relation $\sigma_s\preceq_z\sigma_t$ requires equal memory
metadata and semantic bits. Every concrete result of a typed load from
the target heap must also be possible from the source heap. If $z$ is
present, the restored local environments and frames must agree as well.

The architectural contract $\mathsf{Arch}(s)$ matches service responses
on heaps related by coherent byte families and matching metadata.
Success preserves admissible memory and saved caller frames. It compares
source and target behavior; $\mathsf{Safe}(s)$ compares executions with
different secrets.

For every $s,\mathcal R,\rho,P,P',T,n,f$,
\[
\begin{gathered}
 \mathsf{Arch}(s)\land\Harden_{\mathcal R,\rho}(P)=\mathsf{Some}(P')
 \land\mathsf{Lookup}(n,P)=\mathsf{Some}(f)\\
 \Longrightarrow\ \exists f'.\ \mathsf{Lookup}(n,P')=\mathsf{Some}(f')
 \land\ \forall g,\vec v,\sigma,K,o,\vec e.\
 \mathsf{World}\Longrightarrow\\
 \forall O,\sigma_t,z.\ \Beh(P',K,f',\vec v,o,\vec e,\sigma,O,\sigma_t,z)
 \Longrightarrow\\
 \bigl(\exists\sigma_s.\ \Beh(P,K,f,\vec v,o,\vec e,\sigma,O,\sigma_s,z)
       \land\sigma_s\preceq_z\sigma_t\bigr)
 \ \lor\ \mathsf{SourceUB}(O).
\end{gathered}
\]
The semantic parameters of $\Beh$ are fixed above. The premise
$\mathsf{World}$ requires all of the following:
\begin{itemize}
\item A state outside speculation with empty locals and frames, an
admissible heap, and correctly typed, readable nonvariadic arguments to $f$.
\item An observer defined in $K$ and accepted by inference for $P{+}K$,
with nonvariadic arguments and the source return type prepended when needed.
\item Successful context ABI checking: resolved calls use a supported
checked interface, and callers have valid C-ABI annotations.
\item No context definition overriding $\rho$, valid layouts for $P{+}K$,
address bindings for every global that is present, and $\mathsf{Link}(T,g)$.
\end{itemize}
Finally, $\mathsf{SourceUB}(O)$ means that $O=p\cdot q$ and a source
execution has trace $p\cdot[\mathsf{UB}]$, a failed final state, and no
result: undefined behavior occurs after a prefix shared with the target.

\paragraph{Concrete instance and validation status}
The supplement constructs the hardened \texttt{lookup} program using
Vellvm's finite 64-bit addresses, provenance, layouts, bytes, memory, and
concretization. It starts with empty heaps, null pointers, and $i=n=0$,
so normal execution takes the branch without a load. Rocq compilation
accepts the base initial-state, layout, and linking premises, as well as
the architectural entry, observer, and service-contract premises.
The unused service always reports out-of-memory. The example exercises
no successful load, allocation, declassification, or RNG response.

Independent rechecking fails with a module-typing error that also occurs
when instantiating Vellvm's unchanged concretization factory without
SSLH. The instance therefore remains unvalidated and neither instantiates
all three theorems nor covers a cryptographic core. Checker acceptance
alone does not establish the runtime premises. The generic theorems' assumption lists contain no
admitted lemmas from vendored Vellvm. Their inherited logical principles,
runtime interfaces, and explicit service premises are described in
Appendix~\ref{app:proof-assumptions}.

The proofs share one runtime module to keep memory and event types
consistent. Instruction lemmas are separate from trace composition and
can be reused across programs.

\subsection{The security invariant}
\label{app:sct-invariant}

The proof starts from valid states with equal public inputs and possibly
different secrets. At corresponding control points, the executions have
the same semantic speculation bit. Their current and saved SSA
environments contain related values: Public values agree, Transient
values agree outside speculation, and Secret values may differ.
The heap relation tracks allocations, permissions, pointer provenance,
and public bytes. It also accounts for readability and the possible
concrete outcomes of LLVM values.

The invariant connects the MSF to the semantic speculation state.
A ready flag is all ones during misspeculation. After a branch, the
invariant records the chosen edge until the MSF update completes.
Protection cannot use the old flag in this intermediate phase: it may
not yet reflect the misprediction.

\subsection{Local steps preserve the relation}
\label{app:local-security-proof}

For arithmetic that cannot fail, related Public operands produce equal
results. Public branch conditions give matching observations, and the
invariant records the chosen edge for the next MSF update.

Loads motivate the Transient type. A load from Public memory returns
equal values during normal execution, but misspeculation may redirect it
to secret memory. The resulting values can differ and are typed
Transient. Before an observable use, protection preserves their normal
values or replaces both speculative values with all ones. The pointer
case also accounts for pointer construction and its possible failure.

The store rules preserve the heap relation. Partial operations require
enough operand agreement to match both successful results and faults.
All PHIs read the saved predecessor environment before writing any
destination. Flag PHIs also preserve the flag's origin and pending edge.

At a call, the checked interface and fence protocol establish the
argument and memory relations, preserve the suspended caller, and end
the old flag's lifetime. Return restores related caller frames. External
services use their assumed relational contracts.

\subsection{From local steps to observable executions}
\label{app:trace-composition}

An LLVM instruction may take several semantic steps to read operands,
choose concrete values, or access memory. We match these steps up to the
next observation, allowing different numbers of silent steps. The
matching segments produce the same ordinary observation and preserve
the invariant.

Induction gives a matching execution for every finite observation
prefix. Applying the argument in both directions proves equality of the
possible prefixes under the same attacker strategy. The proof matches
public nondeterministic choices, including allocation addresses; it does
not assume that independent choices agree.

\subsection{Declassification and trusted-service cases}
\label{app:declassification-proof}

The declassification rule requires Public control and an authorized
site, so related executions reach the same site. During normal
execution, each side materializes its operand as a value or a failure.
Every pair of steps with equal outcomes must leave related states.
Unequal outcomes mark the first authorized difference, after matching
all preceding observations.

During misspeculation, declassification returns the operand without
revealing it or changing its speculative type. A value that may differ
between executions still needs protection before an observable use.

Public buffer addresses and lengths ensure that both executions make
the same service request. The assumed contract supplies a matching
response in the other execution: success preserves the state relation,
memory metadata, and saved frames; failure produces the same observation.
Secret output bytes may differ. The service implementation lies outside
the proof.

Induction in both directions gives the trace property of
Appendix~\ref{app:formal-statements}, up to the first authorized difference.

\subsection{Relating source and inserted fragments}
\label{app:insertion-proof}

Architectural preservation compares the original and hardened programs
from the same state and arguments. The source may stay at one instruction
while the target computes masks or updates the MSF. The subsequent
original operation must produce related results and effects.

For example, protection before a load leaves its operand unchanged when
the flag is clean, so both programs access the same address. After the
load, the invariant relates the source destination to the generated SSA
names. The remaining instruction cases follow the same structure,
tracking the CFG edge and the current phase of each inserted sequence.

A decreasing measure bounds the bookkeeping before the next original
operation. Composing the instruction cases restores caller frames and
relates the final observer result. Concrete results and memory satisfy
the refinement relation even when symbolic expression trees differ.
The theorem also allows source undefined behavior after a shared
observation prefix.

\subsection{Proof assumptions}
\label{app:proof-assumptions}

The proofs use Vellvm's address, provenance, layout, concretization, and
byte interfaces. Appendix~\ref{app:formal-statements} gives the input,
memory, linking, context, and service requirements for each theorem.
These requirements constrain the initial state and available operations;
the simulation is proved. Compiler, hardware, and policy limits are
discussed in Section~\ref{sec:limitations}.

The final assumption audits contain no admitted Vellvm lemmas or added
security or preservation axioms. They use Vellvm's classical reasoning,
functional extensionality, and equality principles. The final theorems
were kernel-checked using precompiled dependencies; the full dependency
tree has not been validated by a clean rebuild.

\section{Backend Preservation Obligations}
\label{app:backend-proof}

The Rocq proofs establish security at LLVM IR. Carrying this result to
assembly requires each later pass to preserve the relation below,
including observations from compiler-generated instructions. We consider
the configured SelectionDAG backends for x86-64 Linux and AArch64
Linux/Darwin. Our implementation checks some of these conditions; the
rest are assumptions about LLVM passes, calling conventions, and hardware.

\subsection{Execution domain and required invariant}

The target model observes attempted memory addresses, conditional control
transfers, public results, and failures (Section~\ref{sec:model});
register-only operations are silent. We assume that barriers stop an
existing speculative window and that masks take effect before protected
values are used. These hardware--software contracts~\cite{speculation-contracts}
are assumptions about the target processor.

The initial target states agree on stack bases, argument locations, code
and global addresses, and relevant permissions. Spill and saved-register
storage is separate from source objects, which may occupy other parts of
the frame. We assume enough stack space and source executions without
architectural undefined behavior. Defined nondeterministic operations,
including \texttt{freeze}, require the additional condition below.

At corresponding program points, a proof-only map $J$ locates each live
value in a virtual register, physical register, or private stack slot.
The relation requires:
\begin{enumerate}
\item Live values keep their normal and speculative classifications.
Public observable operands agree, and pointer representations preserve
the correspondence between addresses and provenance.
\item A ready MSF occupies a whole register and equals
$\operatorname{enc}(b)$, where $\operatorname{enc}(\mathsf{false})=0$ and
$\operatorname{enc}(\mathsf{true})=2^{64}-1$. Until an update finishes, a
pending flag retains the old state, captured condition, and chosen edge.
\item An active MSF never resides in memory. Register assignments respect
overlapping live ranges and register aliases. The flag's lifetime ends
at the checked call boundary.
\item Added stack locations have public addresses and are private to the
compiler. Before a reload reaches a Public use, it must be shown Public
by a store/reload check, protected by a ready mask, or covered by a full
barrier.
\end{enumerate}
The flag must also be correct during normal execution: an all-ones mask
would otherwise change an architectural value.

\paragraph{Local matching requirement}
\label{def:backend-unwinding}
Write $q_1\equiv_J q_2$ for this relation, including agreement on control
position, primitive-expansion phase, and semantic speculation bit.
From related configurations, each finite segment ending at the next
observation must have a matching segment with that observation and
related final configurations. This holds in both directions, including
possible failures, except for a first authorized difference at normal
declassification. During primitive expansion, $J$ also tracks temporary
values and the remaining steps.

\paragraph{Conditional implication}
\label{lem:backend-unwinding}
If local matching holds at every reachable phase, induction extends
matching histories one observation at a time. Equal histories select
the same attacker directive. The argument in both directions gives
equality of possible finite prefixes up to authorized releases.
Each compilation stage must establish local matching for its output.

\subsection{A conditional backend guarantee}
\label{thm:backend-preservation}

Suppose the initial calling-convention and loader states satisfy the
invariant, every compiler stage preserves local matching, nondeterministic
choices are resolved safely, and the hardware contracts hold. The two
resulting assembly executions then have matching finite observation
prefixes up to the same authorized declassifications. This includes
accesses added for the stack and constant pool. Preservation of normal
results and memory additionally requires the ordinary passes'
architectural correctness contracts.

The implementation checks the following conditions; each item also
identifies what remains assumed.
\begin{enumerate}
\item \textbf{IR and flag protocol.} The extracted checker recomputes labels
and validates SSA/CFG structure, primitive signatures, and uses of the
current flag. The driver also rechecks serialized IR. Correct transport
through the frontend, extraction, and parser remains trusted.
\item \textbf{MSF and masks.} Register tags and checks enforce the supported
nonspill, call-lifetime, flag, and dependency rules. On x86, a certificate
checks the condition, edge, and live flags used when reusing branch flags.
If reuse cannot be justified early, the scalar MSF update is kept. If a
later check finds that the flags changed or no longer identify the edge,
compilation is rejected. These checks do not reconstruct $J$ for every
ordinary value. The optional AArch64 vector representation uses the same
nonspill and call-lifetime checks. Removing unused MSF operations also
requires final validation; if register allocation invalidates that
removal, the driver can retry while retaining the flag state.
\item \textbf{Ordinary rewrites and value choices.} Disabling select
optimization and x86 CMOV conversion prevents known control-flow changes.
LLVM's machine verifier checks structural invariants. Other rewrites
must still preserve the relation between executions, resolve Public
choices independently of secrets, and preserve the required dependencies.
\item \textbf{Frames and reloads.} The x86 reload checks validate supported
Public store/reload paths or masking patterns. AArch64 uses the stated
barrier or mask policies. Separation of source and compiler storage,
matching argument locations, and surrounding frame transformations
remain assumptions.
\item \textbf{Final machine execution.} Barrier effectiveness, dependencies
at protected uses, matching calling-convention and loader behavior, and
preservation through assembly and linking remain target contracts.
Bounded binary tests provide evidence only for the tested objects and
configurations.
\end{enumerate}

Passes that change instruction boundaries also need a translation of
directives between levels. The two executions at each level use the same
adaptive strategy, but compilation can change the directive sequence.
Added stack observations must match between the compiled executions;
they cannot be erased from the argument.

\subsection{IR preparation, selection and nondeterminism}

Loop simplification, induction-variable rewriting, strength reduction,
constant hoisting, code-generation preparation, and lowering must preserve
values, definedness, Public observable operands, and the association
between captured conditions and CFG edges. New conditionals need an MSF
update or barrier before sensitive uses, even when the condition is Public.
Analyses of dominance, aliases, loops, liveness, and frequency introduce
no instructions and need no execution argument.

Instruction selection and legalization must avoid Secret-indexed tables,
unsupported helper calls, and Secret-dependent branches. New temporaries
receive the join of their operands' classifications. Concrete word
operations preserve the value relation; partial operations must also
preserve possible faults. Passes for atomics, vectors or tiles, garbage
collection, exceptions, thread-local storage, profiling, sanitizers, and
thunks need separate arguments when their features are enabled.

\paragraph{Choosing concrete values safely}
At IR level, the proof can match nondeterministic choices between
executions. When compilation resolves a choice, the chosen Public result
must depend only on matching public state and compilation constants,
or have another justification in the state relation.
For example, \texttt{freeze undef} permits every word in either IR
execution. Lowering it to \texttt{IMPLICIT\_DEF} could reuse a register
containing an old Secret value. The compiled executions could then
choose different observable addresses. Each could still refine its own
IR execution while the pair violates SCT. Requiring defined source
behavior does not rule out this example.

An under-defined value must either have no effect on a Public observation
or be resolved independently of secrets before that use. This condition
also applies to register reuse and partial-register writes. The MSF and
reload checks do not establish it globally. Local rejection of undefined
operands, including restrictions on \texttt{freeze}, is not sufficient.

Selection must also preserve flag and data dependencies of the primitives.
Normal declassification keeps the declared concrete outcome; speculative
declassification keeps the speculative type and produces no disclosure.
If lowering erases a marker, the proof must still track its logical site
and release policy.

\subsection{Machine SSA, CFG transformations and scheduling}

PHI optimization, copy propagation, common-subexpression elimination,
peephole rewrites, dead-instruction removal, and loop motion must preserve
each value's representation. Flag copies retain their origin and ready
or pending phase. CFG rewrites---tail duplication, branch folding,
unreachable-block removal, placement, and relaxation---must preserve the
incoming-edge mapping. Branch inversion changes both the successor
interpretation and the flag update. Edge splitting adds an intermediate
phase; a new join must select its predecessor's flag. The implemented
checks cover supported flag transfers, not all ordinary value rewrites
or concrete-value choices.

Scheduling must preserve initialization, update, and protection
dependencies, together with memory and definedness conditions. A Public
view can leave its input bits unchanged but must preserve these dependencies.
The x86 implementation retains an opaque, tied pseudo until final
certification. Side-effect chains, operand constraints, and instruction
bundles enforce local ordering rules; the scheduling stage must preserve
the full security relation.

\subsection{PHI elimination, coalescing, allocation and calls}

PHI elimination inserts copies on incoming edges. Temporaries used to
resolve cycles must preserve the state relation and remain nonspillable
when they hold an MSF. Two-address conversion, coalescing, allocation,
live-range splitting, and cloning must respect overlapping live ranges,
register aliases, and early-clobber constraints for every live MSF copy.
Unsupported transfers or allocation failure must reject compilation.

Register reuse also needs the concrete-value condition above: the bits
already in a register do not make an undefined value Public. At an
ordinary checked call, the caller ends the flag's lifetime, synchronizes,
and initializes a fresh flag after return. Trusted services use their
separate protocol and contract. Argument moves and saved registers obey
the public calling-convention and frame conditions. Carrying a live flag
across calls would require a different interprocedural proof.

\subsection{Ordinary spills and frame construction}
\label{lem:backend-spills}

A spill adds an address observation, which agrees between executions
when the frame bases and fixed offset agree, even for Secret contents.
Separate compiler storage keeps the spill from changing source memory.
An uncertified reload is Transient when its architectural value is Public,
and Secret otherwise. Its public stack address does not establish the
value's speculative classification.

A ready mask must protect a Transient reload before any Public use,
preserving equal normal values and replacing speculative values with all
ones. Placement and dependencies prevent use of the raw reload; a full
barrier permits only a normal continuation. The x86 store/reload checks
account for reaching full-width stores, overlapping or intervening
writes, aliases, and folded loads. A source annotation is insufficient.
The counter rule accepts recognized immediate increments only when every
incoming path reaches the same constant initializer; it cannot make a
Secret architectural value Public.

The conservative AArch64 profile uses full barriers for ordinary stack
traffic. Its optional masked-reload profile requires a locally live GPR
flag, a mask, and a consumption barrier at the use. Removing duplicate
barriers must preserve coverage of every entry path. Stack and slot
coloring, frame allocation, prologue and epilogue insertion, shrink
wrapping, and frame-index elimination must preserve disjoint live
lifetimes, public stack adjustments, and register saves and restores.
Accesses before MSF initialization require the same frame conditions.

\subsection{Primitive expansion and final rewrites}
\label{lem:backend-primitives}

For concrete words, primitive expansion obeys the following equations.
Initialization combines a target barrier with zeroing the flag. Let $c$
be the Boolean branch condition and $t$ the chosen edge. A true-edge
update computes $c-1$, and a false-edge update computes $-c$, modulo
$2^{64}$. Each is zero exactly when $c=t$ and all ones otherwise. OR with
the old flag therefore represents $b\lor(c\ne t)$. When an update reuses
branch flags, checking must establish that those bits represent this
condition and that other flag uses remain valid. Reasoning about the
whole flag register can miss bits preserved by a partial update.

Protection copies the input and ORs in the mask truncated to its width.
It preserves a concrete normal word and returns $2^w-1$ when the MSF is
poisoned. Dependencies and scratch-register alias conditions must ensure
that the consumer uses this protected result. This word-level equation
does not by itself justify LLVM poison, under-defined values, or pointer
provenance. Initialization uses \texttt{LFENCE} on x86 and \texttt{SB}
or \texttt{DSB SY; ISB} on AArch64. AArch64 uses of masks and updates also
require the stated consumption-barrier contract. Its optional vector
representation must preserve the low-word bits and satisfy the alias,
scratch, nonspill, and lifetime conditions.

After allocation, the obligations also cover copy removal, scheduling,
LEA and addressing rewrites, width and execution-dependency fixes,
branch and layout changes, and load/store pairing. Every component of
a paired access needs a public address. Debug and unwind metadata are
silent; instrumentation that emits instructions needs its own argument.
Final x86 checking covers supported spill and flag patterns. AArch64
expansion runs after branch relaxation. Bundle unpacking, printing,
assembly, relocation, and linking must preserve operands, ordering,
branch targets, and the loader mapping.

\section{Benchmark Preparation and Measurement Details}
\label{app:benchmark-preparation}

Streams, Secretbox, X25519, and SHA-512 include TweetNaCl implementations;
the other routines use fixed versions of libsodium, HACL*, or Kyber.
Some of the 24 operations share code and are not independent algorithm
families. Every compared ChaCha20 wrapper omits the outer context cleanup.

The default comparison uses the same prepared IR, instruction-set
features, drivers, and inputs. Preparation uses inlining, stack-variable
promotion, and loop settings that expose supported scalar instructions.
X25519 keeps its arithmetic \texttt{sel25519} until a later inlining
step. We expand selects with proved-defined operands and Kyber's integer
selects with constant alternatives. Poly1305 retains its branchless
arithmetic; remaining Secret select guards are rejected. Select
optimization and x86 CMOV-to-branch conversion are disabled in all three
versions. VellSSLH also uses the backend restrictions of
Section~\ref{sec:implementation}.

All 24 modules pass VellSCT, so Theorems~\ref{thm:sct}--\ref{thm:release-sct}
apply under their entry, runtime, declassification, and service assumptions,
regardless of the insertion algorithm. Theorem~\ref{thm:preservation}
covers the reference inserter. Fourteen benchmarks use other selective
or read-only variants; ten use a release path without an established
correspondence to the proved source version. Preparation and Public-view
rewrites are also outside that theorem. Functional tests check normal
behavior but do not establish architectural preservation for these variants.

The nine matched trials run on CPU~2, identified by CPUID as a performance
core. We randomize code padding, link order, and measurement order while
keeping the cryptographic objects and driver sources unchanged.
Timing includes the common test loop and output checksum, but excludes
calibration, warmup, reference computations, and full-output checks.
Comparisons with unprotected code and LLVM SLH use the geometric mean
of within-trial ratios. Ultimate SLH was measured separately and uses
ratios of medians. The percentages can therefore differ from ratios of
rounded table entries. Confidence intervals resample whole trials,
including layout variation; no trials are discarded.

\section{Matched Backend Configuration Controls}
\label{app:backend-tuning-controls}

Table~\ref{tab:backend-tuning-controls} compares unprotected code and LLVM
SLH with loads separated from their consuming instructions. For hashes,
it also compares reserving and releasing the reload scratch register.
The input IR and paired, randomized-layout trials match
Table~\ref{tab:backend-tuning}. These controls separate the effects of
ordinary memory-instruction choices from protection costs.

\begin{table}[!htbp]
\centering\small
\setlength{\tabcolsep}{3pt}
\begin{tabular}{lrrrr}
\toprule
Operation & \shortstack{Bare: no fusion\\vs normal (\%)} & \shortstack{SLH: no fusion\\vs normal (\%)} & \shortstack{Tuned vs no-fusion\\SLH (\%)} & \shortstack{Scratch off vs on\\within no fusion\\(\%)} \\
\midrule
SHA3-256 & \shortstack{$-0.53$\\$[-0.87,-0.17]$} & \shortstack{$+0.01$\\$[-0.23,+0.26]$} & \shortstack{$-0.74$\\$[-1.07,-0.40]$} & \shortstack{$-2.01$\\$[-2.52,-1.62]$} \\
SHA3-512 & \shortstack{$+0.21$\\$[-0.02,+0.44]$} & \shortstack{$+0.59$\\$[+0.44,+0.74]$} & \shortstack{$-0.21$\\$[-0.50,+0.11]$} & \shortstack{$-1.17$\\$[-1.46,-0.81]$} \\
SHAKE128 & \shortstack{$-0.14$\\$[-0.39,+0.09]$} & \shortstack{$-0.09$\\$[-0.36,+0.18]$} & \shortstack{$+0.46$\\$[+0.09,+0.78]$} & \shortstack{$-0.28$\\$[-0.63,+0.04]$} \\
SHAKE256 & \shortstack{$+0.99$\\$[+0.52,+1.49]$} & \shortstack{$+0.35$\\$[-0.06,+0.73]$} & \shortstack{$+0.61$\\$[+0.09,+1.18]$} & \shortstack{$-0.39$\\$[-0.82,+0.07]$} \\
Kyber512 keygen & \shortstack{$+1.80$\\$[+1.42,+2.20]$} & \shortstack{$+1.79$\\$[+1.50,+2.13]$} & \shortstack{$+0.19$\\$[-0.02,+0.45]$} & -- \\
Kyber768 keygen & \shortstack{$+2.03$\\$[+1.79,+2.25]$} & \shortstack{$+0.45$\\$[+0.27,+0.63]$} & \shortstack{$-0.45$\\$[-0.69,-0.21]$} & -- \\
Kyber768 encap. & \shortstack{$+1.67$\\$[+1.51,+1.82]$} & \shortstack{$+1.87$\\$[+1.70,+2.04]$} & \shortstack{$-3.50$\\$[-3.81,-3.16]$} & -- \\
\bottomrule
\end{tabular}
\caption{Backend-option controls: paired percentage changes with 95\% intervals.}
\label{tab:backend-tuning-controls}
\end{table}

\begin{table}[!htbp]
\centering\small
\begin{tabular}{lrrrr}
\toprule
Operation & Instructions & MOVs & Function bytes & INIT \\
\midrule
SHA3-256 & $1,499\to1,566$ & $707\to778$ & $6,227\to6,404$ & 1 \\
SHA3-512 & $1,649\to1,726$ & $801\to882$ & $6,677\to6,944$ & 1 \\
SHAKE128 & $2,169\to2,257$ & $1,012\to1,107$ & $9,380\to9,553$ & 1 \\
SHAKE256 & $2,171\to2,259$ & $1,013\to1,108$ & $9,396\to9,553$ & 1 \\
Kyber512 keygen & $7,940\to8,188$ & $3,830\to4,098$ & $34,917\to35,251$ & 3 \\
Kyber768 keygen & $6,763\to6,978$ & $3,279\to3,511$ & $29,761\to30,026$ & 3 \\
Kyber768 encap. & $10,948\to11,400$ & $5,201\to5,685$ & $48,702\to49,644$ & 2 \\
\bottomrule
\end{tabular}
\caption{Static code counts and function sizes before and after tuning; INIT counts initialization sites.}
\label{tab:backend-tuning-inventory}
\end{table}

For SHA3-256, separating loads while keeping the scratch register changes
runtime by $+0.26\%$ (95\% interval $[-0.002,+0.54]\%$). Releasing that
register in the same configuration improves runtime by 2.01\% (change
$[-2.52,-1.62]\%$). For SHA3-512, the first change is an inconclusive
$+0.07\%$, and releasing the register gives a 1.17\% improvement (change
$[-1.46,-0.81]\%$).

Releasing the register changes allocation and instruction selection as
well as removing masks. For SHA3-512,
the instruction count rises from 1,649 to 1,726 and the MOV count from
801 to 882, despite eliminating late masks. The timings measure these
effects together. All MSF initializations and call protocols remain,
and the strict backend and functional checks pass. These changed
binaries have not received new security queries.

\section{Bounded Binary Analysis Coverage}
\label{app:binary-coverage}

All selected configurations complete for eight binaries; the other 16
remain inconclusive (Table~\ref{tab:binary-coverage}). A successful check
covers the tested configurations and speculation window. The instruction
model supports separately tested, restricted forms of rotates and SHLD,
dead-flag intervals, and zero-extending word-register reads. It rejects
other forms.

\begin{table}[!htbp]
\centering\small
\begin{tabular}{p{.34\linewidth}rp{.52\linewidth}}
\toprule
Binaries & Count & Reason the analysis is inconclusive \\
\midrule
Salsa20/XSalsa20 streams & 2 & Some nonempty-input queries exceed resource limits. Zero-length queries and Salsa20's one-byte query finish. \\
Secretbox encrypt, SHA3/SHAKE, Kyber768 keygen & 6 & The model lacks XMM zeroing and transfers. Kyber also uses unsupported word-width operands. \\
Kyber512 keygen & 1 & The model supports rotates into registers, but this binary rotates a value in memory. \\
SHA-512, X25519 scalar/base & 3 & Carry and borrow handling for ADC/SBB is incomplete. X25519 also uses bit-test flags. \\
Kyber encapsulation/decapsulation & 4 & The model does not fully handle 32-bit SBB. \\
\bottomrule
\end{tabular}
\caption{Instruction-model and resource limits for 16 inconclusive binary checks.}
\label{tab:binary-coverage}
\end{table}

These limits belong to the pinned Spectector version, translator, or
adapter; the IR checker can accept the corresponding scalar program.
Register allocation may introduce \texttt{xorps} and
\texttt{movaps}/\texttt{movups} to zero or copy stack data. Their model
must account for 128-bit contents and memory effects. Similarly,
\texttt{adc}/\texttt{sbb} require width-correct carry, borrow, and flag
semantics, including their effects on later arithmetic and branches.
Kyber512 key generation contains \texttt{rolq \$27,-88(\%rbp)}, a rotate
in an ordinary stack slot; the restricted model accepts only
general-purpose register destinations. This operation does not spill
the MSF.

Unsupported instructions leave binary security undecided. Longer time
limits can help an exhausted search, but cannot supply missing semantics.
The queries fix public inputs and memory regions; they do not model
declassification or trusted services. The seven tuned binaries have not
been rechecked, so the default verdicts and diagnostics do not transfer.

\section{Ultimate SLH Comparison}
\label{app:ultimate-comparison}

We compare all 24 default VellSSLH operations with Ultimate
SLH~\cite{ultimate-slh} on the same Intel core. Ultimate uses LLVM~16,
post-load hardening, and all four published options:
\texttt{sbhAll}, \texttt{fixed}, \texttt{store}, and \texttt{vtInstr}.
We adapt one API spelling to the upstream version without changing the
algorithm. Ten inputs use older attribute spellings, which we check
normalize to the same attributes. The IR,
instruction-set profile, test drivers, and reference inputs are otherwise
the same. Upstream LLVM lacks our project-specific load-narrowing option.

\paragraph{Scope of the additional protections.}
Ultimate SLH also protects variable-time instructions, a channel our
observation model does not explicitly cover. None of the 24 measured
cryptographic objects contains integer division, floating-point arithmetic,
or \texttt{REP} instructions; shared runtime-library code is excluded
from this claim. Recompiling every input with the same Ultimate compiler
and settings, changing only \texttt{vtInstr} to false, produces 24
byte-identical ELF objects. The option adds no code in these workloads
and remains enabled in all reported timings.

Both tools protect branch conditions and memory addresses. The
variable-time option adds no code here, but the tools still differ in
their security models and supported programs.

\begin{table}[!htbp]
\centering\small
\setlength{\tabcolsep}{4pt}
\begin{tabular}{lrrrrr}
\toprule
Operation & Bytes & Ultimate & \shortstack{VellSSLH\\default} & \shortstack{$\Delta$ vs\\Ultimate (\%)} & \shortstack{Ratio\\95\% interval} \\
\midrule
Poly1305 & 16,384 & 8.452 & 7.775 & -8.01 & $[0.9165,0.9279]$ \\
SipHash-2-4 & 16,384 & 5.422 & 5.373 & -0.91 & $[0.9875,0.9944]$ \\
SipHash128 & 16,384 & 5.429 & 5.325 & -1.91 & $[0.9798,0.9841]$ \\
ChaCha20 & 16,384 & 24.206 & 23.543 & -2.74 & $[0.9639,0.9774]$ \\
Salsa20 & 16,384 & 67.620 & 45.689 & -32.43 & $[0.6700,0.6843]$ \\
XSalsa20 & 16,384 & 72.440 & 44.612 & -38.42 & $[0.6134,0.6267]$ \\
Secretbox encrypt & 16,384 & 819.762 & 362.844 & -55.74 & $[0.4414,0.4435]$ \\
Secretbox open & 16,384 & 535.922 & 223.031 & -58.38 & $[0.4126,0.4296]$ \\
SHA3-256 & 16,384 & 27.707 & 26.497 & -4.37 & $[0.9521,0.9589]$ \\
SHA3-512 & 16,384 & 50.492 & 49.184 & -2.59 & $[0.9728,0.9762]$ \\
SHA-512 & 16,384 & 42.784 & 33.644 & -21.36 & $[0.7850,0.7892]$ \\
SHAKE128 & 16,384 & 22.982 & 21.750 & -5.36 & $[0.9415,0.9544]$ \\
SHAKE256 & 16,384 & 28.291 & 26.664 & -5.75 & $[0.9415,0.9488]$ \\
X25519 scalar & 32 & 844.987 & 600.199 & -28.97 & $[0.7082,0.7127]$ \\
X25519 base & 32 & 844.236 & 601.333 & -28.77 & $[0.7109,0.7147]$ \\
Kyber512 keygen & -- & 23.531 & 18.869 & -19.81 & $[0.8005,0.8048]$ \\
Kyber512 encap. & -- & 28.501 & 22.351 & -21.58 & $[0.7781,0.7887]$ \\
Kyber512 decap. & -- & 33.572 & 24.382 & -27.38 & $[0.7218,0.7320]$ \\
Kyber768 keygen & -- & 40.697 & 31.159 & -23.44 & $[0.7636,0.7700]$ \\
Kyber768 encap. & -- & 46.375 & 35.868 & -22.66 & $[0.7701,0.7766]$ \\
Kyber768 decap. & -- & 52.823 & 39.374 & -25.46 & $[0.7405,0.7490]$ \\
Salsa20 core & 16 & 0.084 & 0.087 & +3.25 & $[1.0234,1.0370]$ \\
HSalsa20 core & 16 & 0.078 & 0.084 & +7.73 & $[1.0718,1.0814]$ \\
HChaCha20 core & 16 & 0.083 & 0.088 & +6.09 & $[1.0525,1.0639]$ \\
\bottomrule
\end{tabular}
\caption{Ultimate SLH comparison ($\mu$s/call); ratio intervals independently resample trials from each campaign.}
\label{tab:ultimate-comparison}
\end{table}

Ultimate uses nine fixed-layout trials; VellSSLH uses nine
randomized-layout trials. Confidence intervals resample each set's trial
medians independently. The geometric mean of the runtime ratios is
0.80349; 21 intervals are below one, while the three cores remain slower.
The different compilers and measurement procedures prevent attributing
the result solely to hardening policy.

Functional tests cover 19,812 cases per Ultimate or control version.
Kyber has consistency tests but no independent known-answer validation.
The audit covers 24 linked routines, 1,944 timed batches, and 2,555 logs.
Ultimate has not been checked with Spectector, and this comparison
excludes the separately tuned binaries.

\end{document}